\documentclass[12pt]{article}

\usepackage{latexsym}
\usepackage{amsmath}
\usepackage{amsfonts}
\usepackage{amssymb}
\usepackage{psfrag}
\usepackage{graphicx}
\usepackage{xcolor}

\newcommand{\comment}[1]{}

\usepackage{natbib}
\setcitestyle{authoryear,open={(},close={)}}

\renewcommand{\theequation}{\arabic{equation}}

\newtheorem{theorem}{Theorem}[section]

\newfont{\cmrten}{cmr10}

\newcommand{\imathbf}[1]{\mbox{\boldmath $#1$}}

\newcommand{\ixi}{\imathbf{\xi}}
\newcommand{\iSigma}{\imathbf{\Sigma}}
\newcommand{\iOmega}{\imathbf{\Omega}}

\newcommand{\iap}{\boldsymbol{\alpha}}
\newcommand{\ib}{\boldsymbol{\beta}}
\newcommand{\isg}{\widehat{\sigma}}

\newcommand{\ig}{\boldsymbol{\gamma}}
\newcommand{\iZ}{Z}

\newcommand{\ia}{\widehat{\alpha}}

\outer\def\endcp{\par\ifdim\lastskip<\medskipamount\removelastskip
  \penalty 55 \fi\medskip\rm}

\outer\def\endpro{\par\ifdim\lastskip<\medskipamount\removelastskip
  \penalty 55 \fi\medskip\rm}
\def\sqr#1#2{{\quad\vbox{\hrule height.#2pt
        \hbox{\vrule width.#2pt height#1pt \kern#1pt
             \vrule width.#2pt}
                     \hrule height.#2pt}}}

\def\ur#1{\uppercase\expandafter{\romannumeral#1}}

\newcommand{\RR}{\mathbb {R}}

\def\gop{{\buildrel P \over \longrightarrow}}
\def\god{{\buildrel {\cal D} \over \longrightarrow}}

\def\hat{\widehat}

\definecolor{purple}{rgb}{0.50,0.00,0.50}
\definecolor{orange}{rgb}{0.80,0.52,0.00}
\def\colb{\color{blue}}
\def\colr{\color{red}}

\def\m{\medskip}
\def\b{\bigskip}

\begin{document}

\title{Estimation and Hypothesis Testing of Strain-Specific Vaccine Efficacy
with Missing Strain Types with Applications to a COVID-19 Vaccine Trial}
\author{
Fei Heng$^{1}$ ,Yanqing Sun$^{2,*}$, and Peter B.\ Gilbert$^{3,4}$\\
$^{1}$ University of North Florida, \\
$^{2}$ University of North Carolina at Charlotte, \\
$^{3}$ University of Washington, and \\
$^{4}$ Fred Hutchinson Cancer Research Center, U.S.A.\\
$^*$E-mail: yasun@uncc.edu
}

\date{}
\maketitle

%%%
\comment{
\author{
Fei Heng$^{1}$ ,Yanqing Sun$^{2,*}$, and Peter B.\ Gilbert$^{3,4}$\\
$^{1}$ Department of Mathematics, \\University of North Florida, Jacksonville, U.S.A.\\
$^{2}$ Department of Mathematics and Statistics, \\University of North Carolina at Charlotte, Charlotte, NC 28223, U.S.A.\\
$^{3}$ Department of Biostatistics, \\University of Washington, Seattle, WA 98195, U.S.A.\\
$^{4}$ Vaccine and Infectious Disease and Public Health Sciences Divisions, \\
Fred Hutchinson Cancer Research Center, Seattle, WA 98109, U.S.A.\\
$^*$E-mail: yasun@uncc.edu
}
}
%%%

\begin{abstract}
Statistical methods are developed for analysis of clinical and virus genetics data from phase 3 randomized,
placebo-controlled trials of vaccines against novel coronavirus COVID-19.
Vaccine efficacy (VE) of a vaccine to prevent COVID-19 caused by one of finitely many genetic strains of SARS-CoV-2 
may vary by strain.
 The problem of assessing differential VE by viral genetics can be formulated under a competing risks model where the 
endpoint is virologically confirmed COVID-19 and the cause-of-failure is the infecting SARS-CoV-2 genotype. 
Strain-specific VE is defined as one minus the cause-specific hazard ratio (vaccine/placebo). 
For the COVID-19 VE trials, the time to COVID-19 is right-censored, and a substantial percentage of 
failure cases are missing the infecting virus genotype.
We develop estimation and hypothesis testing procedures for strain-specific VE when the 
failure time is subject to right censoring 
and the cause-of-failure is subject to missingness, focusing on $J \ge 2$ discrete categorical unordered or ordered virus genotypes.
 The stratified Cox proportional hazards model is used to relate the cause-specific outcomes to explanatory variables. 
The inverse probability weighted complete-case (IPW) estimator
 and the augmented inverse probability weighted complete-case (AIPW) estimator are investigated.
Hypothesis tests are developed to assess whether the vaccine provides at least a specified level of efficacy against some 
viral genotypes and whether VE varies across genotypes, adjusting for covariates. The
finite-sample properties of the proposed tests are studied through simulations and are shown to have good performances.
In preparation for the real data analyses, the developed methods are applied to a pseudo dataset mimicking the Moderna COVE trial. 
%The developed methods are applied to the data from Moderna COVID-19 vaccine trial. 
\end{abstract}


\section{Introduction}

Randomized, placebo-controlled vaccine efficacy (VE) trials have demonstrated that several SARS-CoV-2 candidate vaccines 
prevent acquisition of COVID-19 with VE level above 50\% and reaching up to 95\% 
(e.g., Polack et al., 2020; Baden et al., 2020)\nocite{Badenetal2020,Polacketal2020}. All of these vaccines
 use the so-called Wuhan or Washington strain (henceforth WA strain) of SARS-CoV-2 in the vaccine construct.
Genetic variability of SARS-CoV-2 viruses has been increasing over time, with several variants emerging that
have several genetic mutations compared to the WA strain, generating concern that
the level of VE could be lower against certain variants (Lauring and Hodcroft, 2021).\nocite{lauring2021genetic} 
For efficacy results reported through about February 2021, the viruses circulating during the trial were 
almost all WA strains or very slight variants (1 or 2 mismatches), such that the trials did not provide
information on VE against variants.
For recent efficacy results for trials in the United Kingdom, Brazil, and South Africa, variants dominated the circulating
strains, such that estimates did assess VE against variants, and the
estimates were lower than for trials
in regions where WA strains dominated (Madhi et al., 2021; Sadoff et al., 2021).\nocite{Madhietal2021,Sadoffetal2021}
To date the published statistical analyses for understanding how VE may depend on variants has
consisted of analyses of VE against
all SARS-CoV-2 strains circulating in a given geographic region, for example the ENSEMBLE trial of the
 Ad26.COV2.S vaccine reported
an estimate of VE of 66.9\% in the U.S. where the WA strain predominated and an estimate of VE
of 52.0\% in South Africa where the B.1.351 variant strain predominated (Sadoff et al., 2021).
%Results have not been reported on how VE depends on variants within a trial or region, where comparing efficacy by 
%variant for the same randomized study population would enable inferences about a real biological difference in VE.

Many of the COVID-19 VE trials are sequencing the SARS-CoV-2 {\it spike} gene for all 
COVID-19 primary endpoint cases, 
where in general the vaccines only include the {\it spike} gene.  These data will enable sieve analysis 
(Gilbert, Ashby, Self, 1998; Edlefsen et al., 2015; Neafsey et al., 2015; Juraska et al., 2018),\nocite{gilbert1998statistical,Edlefsenetal2015,Neafseyetal2015,Juraskaetal2018} which, based on a competing risks
failure time data set-up, assesses whether and how VE depends on genetic features of the pathogen strain causing the disease endpoint.
Rolland and Gilbert (2021)\nocite{RollandGilbert2021} briefly discussed motivation and 
applications of sieve analysis in SARS-CoV-2 VE trials. 
While many statistical methods of sieve analysis have been developed, some features of the forthcoming data sets for the COVID-19 VE trials
require some novel methods development.  First, the methodology needs to allow for missing sequences, because sequencing technology is only able
to measure the {\it spike} sequence if the viral load of the sample used for sequencing is sufficiently high (Xiao et al., 2022).\nocite{xiao2020multiple} 
 A wealth of data in natural history studies suggest an expected 20-30\% of placebo arm COVID-19 endpoint cases
will have missing sequences, and the rate of missingness will generally be higher in vaccine arm COVID-19 endpoint cases, 
given that the vaccines usually have some impact to
suppress viral load.  Second, the methodology needs to handle $J$ discrete categorical genotypes with $J>2$ and allowing the multiple genotypes to be either 
unordered categorical or ordered
categorical (e.g., a Hamming distance that is the number of amino acids in the Spike protein that are mismatched to the WA vaccine strain).
Third, 
it is useful for the methods to provide inferences for whether VE differs across genotypes (with variation termed a ``sieve effect"), 
not only providing
separate inferences about VE
against each individual genotype. 
Lastly, the methodology needs to accommodate that the background incidence of COVID-19 can vary over calendar time, given waxing and waning outbreaks.
In this work, we focus on addressing these needs through a proportional hazards model, which is reasonable for 
the COVID-19 VE trials during their primary periods of follow-up, given that these periods last less than 6 months and immune 
responses induced by the vaccines are fairly stable during these periods.  This indicates that the Cox model assumption of
time-constant VE against any given genotype 
is a reasonable assumption, at least approximately.

Among Cox model methods that handle missing causes of failure,
Goetghebeur and Ryan (1995) used weighted Cox modeling, and Lu and Tsiatis (2001) used a parametric model
for the probability of observing a sequence and used multiple imputation to predict missing genotypes
from auxiliary covariates.  Adapting the theory of Robins, Rotnitzky, and Zhao (1994),\nocite{RobinsRotnitzkyZhao1994} Gao and Tsiatis (2005) 
considered linear transformation 
models -- with the Cox model a special case -- developing inverse probability weighted (IPW) and
augmented IPW (AIPW) methods; Hyun et al. (2012) also developed IPW and AIPW methods for the Cox model.
These papers focused on two causes of failure, and did not provide techniques for sieve effect tests.  Moreover, auxiliary covariates were considered,
but for the COVID-19 application auxiliary marks are more valuable. (A `mark' is a random variable only meaningfully defined in failure endpoint cases.)  
In particular, the key auxiliary mark is the SARS-CoV-2 viral load from the blood sample used for sequencing the virus.  In addition, for the applications
it is useful to allow separate baseline hazards for different calendar intervals of enrollment, as one way to handle unpredictable secular trends
in placebo COVID-19 arm incidence.  While all of the methods could be devised to allow multiple baseline hazards, 
the available implementations typically do not include this implementation.  This current work most closely resembles that of Hyun et al. (2012), 
where we also develop IPW and AIPW methods,
and take on the new features not considered previously of handling $J>2$ unordered or ordered categorical genotypes, and 
developing hypothesis testing procedures for multiple new questions of interest including the assessment of sieve effects.  The methods are implemented
in the R package {\it cmprskPH} available at 
https:$\slash \slash$github.com$\slash$fei-heng$\slash$cmprskPH. 

%{\colr 
The rest of this article is organized as follows. 
In Section \ref{Preliminaries}, we present the mathematical framework for estimating VE against specific genotypes which are subject to missingness.
% stratified cause-specific Cox model
%Assumptions and some basic notations are introduced.
Section \ref{estimation} is devoted to demonstrating the development of IPW and AIPW estimation methods. 
Asymptotic properties for the proposed estimators are established in Section \ref{asymptotic}, with proofs in the Web Appendix A. 
In Section \ref{inference}, the confidence intervals and hypothesis testing procedures for VE are derived. 
We conduct simulation studies to examine the finite-sample performance of  estimators and the tests in Section \ref{simulation},
and we apply our methods to a pseudo dataset mimicking the Moderna COVE trial in Section \ref{application}.
%}

\section{Stratified cause-specific proportional hazards models and missing causes}
\label{Preliminaries}

Let $\tau$ be the duration of the vaccine trial. Let $T$ be the failure time, $V$ the cause of failure (also termed as discrete mark or type of infecting strain in VE trials), 
and $Z(t)$ a possibly time-dependent $p$-dimensional covariate.  
Statistical interest focuses on the conditional cause-specific hazard rate of cause $j$ defined by
\[
  \lambda_{j}(t|z(t))=\lim_{\Delta t\downarrow
  0}\frac{1}{\Delta t} P\bigl( t\leq T < t+\Delta t, V=j|T
  \geq t, \iZ(t)=z(t) \bigr ),
\]
for $j=1,\ldots,J$. The function $\lambda_j(t|z(t))$ is the instantaneous
failure rate from cause $j$ at time $t$ in the presence of the
other failure types.
For VE trials, we specifically consider the covariate $z(t)=(z_1,z_2(t))^{T}$,
where $z_1$ is the treatment group indicator (1=vaccine; 0=placebo) and $z_2(t)$
the vector of other covariates. 
Vaccine efficacy to reduce susceptibility to strain $j$  at time $t$ is defined as
\[
VE_j(t |z_2(t)
)=1-\dfrac{\lambda_j(t|z_1=1,z_2(t))}{\lambda_j(t|z_1=0,z_2(t))}.
\]
Gilbert (2000) discussed the assumptions required for the strain-specific vaccine efficacy to have a meaningful biological interpretation. \nocite{Gilbert2000a}
\comment{\colr Under the assumption of an equal distribution,
conditional on the covariate $z_2$, of exposure to each
strain $j\in\{1,\ldots,J\}$ during the follow-up period $[0,\tau]$
for vaccine and placebo recipients (justified by randomization and
blinding), $VE_j(t |z_2 )$ approximately equals the
relative multiplicative reduction in susceptibility to strain $j$
for vaccine versus placebo recipients conditional on
$z_2$ under a fixed amount of exposure to strain $j$ at
time $t$.}

In practice, different key subgroups (e.g., men and women; individuals living in different
geographic regions, individuals enrolled during different calendar intervals) typically have different baseline cause-specific hazards of failure.
The stratified cause-specific  proportional hazard regression model postulates that the conditional cause-specific hazard function 
for cause $j$  for an individual in the $k$th stratum with the covariate value
$z(t)=(z_1,z_2^T(t))^{T}$ equals
\begin{equation}\label{causeM1}
  \lambda_{kj}(t|z(t))=\lambda_{0kj}(t)\exp(\ib_j^{T}z(t))
    =\lambda_{0kj}(t)\exp(\alpha_{j}z_{1}+
  \ig_j^{T}z_2(t)), 
\end{equation}
for $j=1,\ldots,J$ and $k=1,\ldots, K$, where $\lambda_{0kj}(\cdot)$ is an unspecified cause-specific baseline hazard
function for the $k$th stratum and $K$ is the number of strata.
Here $\ib_{j}=(\alpha_{j}, \ig_j)$ is a $p$-vector of
regression parameters for the $j$th strain. Define  $\ib=(\ib_1,\ldots,\ib_J)$.
Model (\ref{causeM1}) allows different baseline functions for different strata.
Similar generalizations of the Cox model were studied by Dabrowska (1997). \nocite{Dabrowska1997}
%{\colr 
Under model (\ref{causeM1}), the covariate-adjusted strain-specific vaccine efficacy $VE_j$ is one minus $\exp(\alpha_j)$.
%}

\comment{
In the special case of
a single group variable, ($z=1$ indicates the treatment group,
$z=0$ the placebo group), model (\ref{causeM1}) reduces to
\[
  \lambda_{kj}(t|z=0)=\lambda_{0kj}(t) \quad \text{and} \quad
  \lambda_{kj}(t|z=1)=\lambda_{0kj}(t)\exp(\alpha_j).
\]
Thus in the vaccine trial $\alpha_j$ is the log-relative risk 
among vaccine vs. placebo recipiens of disease caused by strain $j$.
}
%\section{Stratified cause-specific proportional hazards models and missing causes}
%\label{Preliminaries}

%{\colr 
To ease the notation in the method development, we generally use $Z(t)=(Z_1,Z_2^T(t))^{T}$ 
to represent the covariate process on $[0,\tau]$, denoted by $Z(\cdot)$.
%Following the standard notation in the survival analysis literature, 
The right-censored failure time data are observations of  $(X,\delta,Z(\cdot))$, where $X=\min\{T,C\}$, $\delta=I(T\le C)$, and $C$ is a censoring random variable.
%Let $V$ be the cause of failure, $V \in \lbrace 0, \cdots, J \rbrace$. 
In a competing risks framework, the cause $V$ can only be observed when failure occurs, whereas it is unknown if the failure time $T$ is censored. 
Then, the completely observed right-censored competing risks data are observations of the random variables $(X, Z(\cdot), V)$ for $\delta=1$ 
and $(X, Z(\cdot))$ for $\delta=0$.
%}

%$\clubsuit$ Peter: A problem in notation has developed?  Above you used $Z(t)=(Z_1,Z_2(t))^{T}$, yet now you're only using
%$Z(t)$.  Clarify or use more consistent notation?  $\clubsuit$
%
%{\colb Fei: add ``Let $\tau$ be the duration of the study. To ease the notation in the method development, we generally use $Z(t)=(Z_1,Z_2^T(t))^{T}$ 
%to represent the covariate process on $[0,\tau]$, denoted by $Z(\cdot)$" at the beginning in the previous paragraph to clarity the usage of $Z(t)$ and $Z(\cdot)$}

%Next we introduce some notation and assumptions that are used throughout the article. 

%{\colr 
In the presence of missing causes, we introduce a binary indicator, $R$, representing whether all possible data are observed for a subject.
$R$ equals to one if either $\delta = 0$ (right-censored) or if $\delta=1$ and $V$ is observed, and zero otherwise. 
The auxiliary covariates $A$ may be helpful for predicting missingness and for informing about the distribution of missing causes.
$A$ may include useful auxiliary marks because causes can only be missing for failures. 
In the COVID-19 application presented in Section \ref{application}, 
$A$  is the SARS-CoV-2 viral load measured in COVID-19 endpoint cases, a continuous mark that is possibly associated with the probability of missingness and the strain type $V$.
%}

\comment{
$R=1$ if either $\delta = 0$ (right-censored) or if $\delta=1$ and $V$ is observed; and $R=0$ otherwise. Auxiliary variables $A$ may be helpful for
predicting missing causes. Since the cause can only be missing for failures, supplemental information is potentially useful only for failures, 
for predicting missingness and for informing about the distribution of missing causes.
The relationship between $A$ and $V$ can be modelled to help predict $V$.
}

We assume that the censoring time $C$ is conditionally independent
of $(T,V)$ given $Z(\cdot)$ for an individual in the $k$th stratum.
We also assume the cause $V$ is
missing at random (Rubin, 1976); \nocite{Rubin1976}
that is, given $\delta=1$ and
$W=(T,Z(T),A)$ of an individual in the $k$th stratum, the probability that the cause
$V$ is missing depends only on the observed $W$, not on the
value of $V$; this assumption is expressed as
\begin{equation}
\label{mar1} r_k(W) \equiv
P(R=1|\delta=1,W)=P(R=1|V,\delta=1,W).
\end{equation}
Let $\pi_k(Q)=P(R=1|Q)$ where
$Q=(\delta,W)$. Then $\pi_k(Q)=\delta r_k(W)+(1-\delta)$.
The missing at random assumption (\ref{mar1}) also implies that
$V$ is independent of $R$ given $Q$:
\begin{equation}
\label{mar3} 
\rho_{kj}(W) \equiv P(V=j |\delta=1,W)= P(V=j |R=1,\delta=1, W).
\end{equation}
For an observed value $w$ of $W$ of an individual in the $k$th stratum, we write $r_k(w)=P(R=1|\delta=1,W=w)$
and $\rho_{kj}(w)=P(V=j |\delta=1,W=w)$.
The stratum-specific definitions of $r_k(w)$ and $\rho_{kj}(w)$
leave the options for the models of the probability of complete-case and cause distribution
to be different for different strata.

Let $n_k$ be the number of subjects in the $k$th stratum; the total sample size is $n=\sum_{k=1}^K n_k$. 
Let $\{X_{ki},Z_{ki}(\cdot),\delta_{ki},R_{ki},V_{ki},A_{ki}; i=1,\ldots, n_k\}$
be iid replicates of $\{X,Z(\cdot),\delta,$ $R,V,A\}$ from the $k$th stratum.
The observed data are denoted by $\{O_{ki}; i=1\ldots, n_k, k=1,\ldots, K\}$,
where
 $O_{ki}=\{X_{ki},Z_{ki}(\cdot),R_{ki},R_{ki}V_{ki},A_{ki}\}$ for $\delta_{ki}=1$
and $O_{ki}=\{X_{ki},Z_{ki}(\cdot),R_{ki}=1\}$ for $\delta_{ki}=0$.
We assume that $\{O_{ki}; i=1\ldots, n_k, k=1,\ldots, K\}$ are independent for all subjects.
Similarly, we denote
$W_{ki}=(T_{ki},Z_{ki}(T_{ki}), A_{ki})$ and $Q_{ki}=(\delta_{ki},W_{ki})$.

We consider a parametric model $r_k(W_{ki},\psi_k)$ for $r_k(W_{ki})$, where $\psi_k$ is an unknown 
vector of parameters to be further discussed in the next section. 
Let $\pi_k(Q_{ki},\psi_k)=\delta_{ki}r_k(W_{ki},\psi_k)+(1-\delta_{ki})$.
Additional notation is introduced in the following. For $\ib\in
\RR^p$, $t\ge 0$, let $Y_{ki}(t)=I(X_{ki}\ge t)$,
$$S_k^{(j)}(t, \ib)=n_k^{-1} \sum_{i=1}^{n_k} Y_{ki}(t)
 \exp \{\ib^T  Z_{ki}(t)\}  Z_{ki}(t) ^{\otimes j},$$
$$\tilde S_k^{(j)}(t, \ib,\psi_k)=n_k^{-1} \sum_{i=1}^{n_k} R_{ki} (\pi_k(Q_{ki},\psi_k))^{-1} Y_{ki}(t)
 \exp \{\ib^T  Z_{ki}(t)\}  Z_{ki}(t) ^{\otimes j},$$
 where for any $z\in \RR^p$,  $z^{\otimes 0}=1$,
 $z^{\otimes 1}=z$ and $z^{\otimes 2}=zz^T$.
Define
 $s_k^{(j)}(t, \ib)=ES_k^{(j)}(t, \ib)$ and
 $\tilde s_k^{(j)}(t, \ib,\psi_k)=E\tilde S_k^{(j)}(t, \ib,\psi_k)$.
Under the missing at random assumption (\ref{mar1}), $s_k^{(j)}(t,
\ib)=\tilde s_k^{(j)}(t, \ib,\psi_k)$ if the model
$r_k(W_{ki},\psi_k)$ is correctly specified. Let
\begin{eqnarray*}
J_{k} (t,\ib)& =& \frac{S_k^{(2)}(t,\ib)}{S_k^{(0)}(t,\ib)} -
\Big  (\frac{S_k^{(1)}(t,\ib)}{S_k^{(0)}(t,\ib)}\Big  )^{\otimes
2}, \\
\tilde J_{k} (t,\ib,\psi_k)& =& \frac{\tilde
S_k^{(2)}(t,\ib,\psi_k)}{\tilde S_k^{(0)}(t,\ib,\psi_k)} - \Big 
(\frac{\tilde S_k^{(1)}(t,\ib,\psi_k)}{\tilde
S_k^{(0)}(t,\ib,\psi_k)}\Big  )^{\otimes 2},
\end{eqnarray*}
$$\bar Z_k(t,\ib)=\frac{S_k^{(1)}(t,\ib)}{S_k^{(0)}(t,\ib)},
\quad \tilde Z_k(t,\ib,\psi_k)=\frac{\tilde
S_k^{(1)}(t,\ib,\psi_k)}{\tilde S_k^{(0)}(t,\ib,\psi_k)}.$$ Let
$\bar z_k(t,\ib)={s_k^{(1)}(t,\ib)}/{s_k^{(0)}(t,\ib)}$ and
$I_k(t,\ib) = {s_k^{(2)}(t,\ib)}/{s_k^{(0)}(t,\ib)} - (\bar
z_k(t,\ib) )^{\otimes 2}$.

\section{Estimation procedures}
\label{estimation}

When there are no missing causes, for each $j$, $\ib_j$ in model
(\ref{causeM1}) can be estimated by maximizing the local
log-partial likelihood function % (Kalbfleisch and Prentice (2002)): \nocite{albfleischPrentice2002}
\begin{equation}
\label{loglik} l(j,\ib_j)=\sum_{k=1}^K\sum_{i=1}^{n_k}
 \int_0^\tau  \Big  [\ib_j^T Z_{ki}(t)-\log\Big 
(\sum_{j=1}^{n_k} Y_{kj}(t)e^{\ib_j^TZ_{kj}(t)}\Big )\Big ]\, N_{kij}( dt),
\end{equation}
where $N_{kij}(dt)=I(X_{ki}\le t,
\delta_{ki}=1, V_{ki}=j)$ is the counting process with a
jump at the uncensored failure time $X_{ki}$ and the associated cause $V_{ki}=j$. 
Taking the derivative of $ l(j,\ib)$ with respect to $\ib$ gives the score function
\begin{equation}
\label{score}
U(j,\ib_j)=\sum_{k=1}^K\sum_{i=1}^{n_k} \int_0^\tau \Big  (Z_{ki}(t)-\frac{S_k^{(1)}(t,
\ib_j)}{S_k^{(0)}(t, \ib_j)}\Big )\, N_{kij}(dt).
\end{equation}
The maximum partial likelihood estimator is a solution to $U(j,\ib_j)=0$.

\subsection{Inverse probability weighted complete-case estimation}
\label{ipw}

Following Horvitz and Thompson (1952), inverse probability weighting of complete-cases
has been commonly used in missing data problems. \nocite{HorvitzThompson1952}
Let $r_k(W_{ki},\psi_k)$ be the parametric model for the probability of
complete-case $r_k(W_{ki})$ defined in (\ref{mar1}), where $\psi_k$
is a $q$-dimensional parameter. For example, one can assume the
logistic model with ${\rm logit}(r_k(W_{ki},\psi_k))=\psi_k^T
W_{ki}$ for those with $\delta_{ki}=1$.
 By (\ref{mar1}), the
maximum likelihood estimator
$\hat\psi=(\hat\psi_1,\ldots,\hat\psi_K)$ of
$\psi=(\psi_1,\ldots,\psi_K)$ is obtained by maximizing the observed
data likelihood,
\begin{equation}
\label{lik_psi}
\prod_{k,i}\{r_k(W_{ki},\psi_k)\}^{R_{ki}\delta_{ki}}\{1-r_k(W_{ki},\psi_k)\}^{(1-R_{ki})\delta_{ki}}.
\end{equation}

We propose the following inverse probability weighted (IPW) estimating equation for
 $\ib$:
\begin{equation}
\label{score_ipw}
U_{I}(j,\ib_j,\hat\psi)=\sum_{k=1}^K\sum_{i=1}^{n_k} \int_0^\tau   \big (Z_{ki}(t)-\tilde Z_k(t,
\ib_j,\hat\psi_k)\big)\frac{R_{ki}}{\pi_k(Q_{ki},\hat\psi_k)}\,N_{kij}(dt).
\end{equation}
The IPW estimator of $\ib_j$ solves the above equation and is denoted by $\hat\ib_{j,I}$.
%Let $\hat\ib_I=(\hat\ib_{1,I}^T, \ldots, \hat\ib_{J,I}^T)^T$.
\comment{\colr In Web Appendix A.1, we show that $\hat\ib_I$ is consistent for the true value $\ib$ and $n^{1/2}\big(\hat{\ib}_{I}-\ib \big)$ converges in distribution to a mean-zero Gaussian random vector with covariance matrix $\iOmega_I$  if  the model for $r_k(W_{ki})$ is correctly specified under mild regularity conditions.}

Let  $\Lambda_{0kj}(t)=\int_0^t\lambda_{0kj}(s)\,ds$ be the cumulative baseline function for each $k$ and $j$.
Let $K(\cdot)$ be a kernel function with bandwidth $h$ and  $K_{h}(x)=K(x/h)/h$.
The baseline function $\lambda_{0kj}(t)$ can be estimated by
$\hat\lambda_{0kj}^{I}(t)$, obtained by
smoothing the increments of the following estimator of the cumulative baseline function $\Lambda_{0kj}(t)$:
\begin{equation}
\label{baseline_IPW} \hat
\lambda_{0kj}^{I}(t)=\int_0^\tau K_{h}(t-s)\,  \hat\Lambda_{0kj}^{I}(ds), \nonumber
\end{equation}
 where 
\begin{equation}
\label{baselineCUM_IPW} 
\hat\Lambda_{0kj}^{I}(t)=\sum_{i=1}^{n_k}\int_0^t
\frac{R_{ki}}{\pi_k(Q_{ki},\hat\psi_k)} \frac{N_{kij}(ds)}{n_k\tilde S_{k}^{(0)}(s,\hat\ib_{j,I},\hat\psi_k)}. \nonumber
\end{equation}

\subsection{Augmented inverse probability weighted complete-case estimation}
\label{aipw}

The IPW estimator $\widehat\ib_{j,I}$ uses only complete cases and is inefficient. 
To increase estimation efficiency, we propose the augmented inverse probability weighted complete-case (AIPW) estimating function 
following the idea of Robins et al. (1994). \nocite{RobinsRotnitzkyZhao1994}
The proposed AIPW estimating equation utilizes available information for individuals with missing causes through a consistent estimator of 
$\rho_{kj}(W)$, the conditional distribution of the failure cause.

In the case that $\sum_{j=1}^J \rho_{kj}(W_{ki})=1$, we posit parametric models $\rho_{kj}(W_{ki},\varphi_{kj})$ for $\rho_{kj}(W_{ki})$ for $j=1,\ldots, J-1$, where $\varphi_{kj}$ are unknown parameters. It is natural to use a logistic multinomial regression model $\text{logit}\{\rho_{kj}(W_{ki},\varphi_{kj})\}=W_{ki}^T  \varphi_{kj}$ for $j=1,\ldots, J-1$, but  other parametric models can also be accommodated.  
Under the MAR assumption (\ref{mar3}), 
 $\rho_{kj}(W_{ki})$ can be estimated using the complete cases with $R_{ki}=1$ and $\delta_{ki}=1$. 
The maximum likelihood estimator $\widehat{\varphi}_{kj}$ of $\varphi_{kj}$ can be obtained by maximizing the likelihood based on complete-case data
\begin{equation*}
\prod_{k=1}^K \prod_{i=1}^{n_k}  \Big(\prod_{j=1}^{J-1} \{\rho_{kj} (W_{ki},\varphi_{kj}) \}^{I(V_{ki}=j) R_{ki} \delta_{ki} }
\{1-\prod_{j=1}^{J-1} \rho_{kj} (W_{ki},\varphi_{kj}) \}^{I(V_{ki}=J) R_{ki} \delta_{ki} } \Big).
\end{equation*}
Since $\widehat\varphi_{kj}$ is the maximum likelihood estimator, it follows that for a correctly specified model $\rho_{kj}(W_{ki},\varphi_{kj})$, $\widehat \varphi$ consistently estimates $\varphi_{kj}$, the true value of the parametric component model $\rho(W_{ki},\varphi)$.
% (Haberman, 1974, 1977; Gourieroux and Monfort, 1981). %\citep{Ha1,Ha2,GM}.
Denote  $\hat\rho_{kj}(W_{ki}) = \rho_{kj}(W_{ki},\hat\varphi_{kj})$ for $j=1,\ldots, J-1$. Then, $\rho_{kJ}(W_{ki})$ can be consistently estimated by $\hat\rho_{kJ}(W_{ki})=1-\sum_{j=1}^{J-1} \hat\rho_{kj}(W_{ki})$.
Let $\hat\rho(\cdot)=\{\hat\rho_{kj}(W_{ki}), k=1,\ldots, K, j=1,\ldots, J \}$.

\b

\comment{
Let $r_k(W_{ki},\psi_k)$ be the parametric model for the probability of
complete-case $r_k(W_{ki})$ defined in (\ref{mar1}), where $\psi_k$
is a $q$-dimensional parameter. For example, one can assume the
logistic model with ${\rm logit}(r_k(W_{ki},\psi_k))=\psi_k^T
W_{ki}$ for those with $\delta_{ki}=1$.
 By (\ref{mar1}), the
maximum likelihood estimator
$\hat\psi=(\hat\psi_1,\ldots,\hat\psi_K)$ of
$\psi=(\psi_1,\ldots,\psi_K)$ is obtained by maximizing the observed
data likelihood,
\begin{equation}
\label{lik_psi}
\prod_{k=1}^K \prod_{i=1}^{n_k} \{r_k(W_{ki},\psi_k)\}^{R_{ki}\delta_{ki}}\{1-r_k(W_{ki},\psi_k)\}^{(1-R_{ki})\delta_{ki}}.
\end{equation}
}

Let $N^{x}_{ki}(t)=I(X_{ki}\le t,\delta_{ki}=1)$,
$N^v_{ki}(v)=I(V_{ki}\le v)$. Following Robins {\sl et al.}\ (1994), \nocite{RobinsRotnitzkyZhao1994}
we obtain the following AIPW
estimating equation for $\ib$:
\begin{eqnarray}
\label{score_aI} & &U_{A}(j,\ib_j,\hat\psi,\hat\rho(\cdot))
=\sum_{k=1}^K\sum_{i=1}^{n_k}  \int_0^\tau  \big(Z_{ki}(t)-\bar Z_k(t,\ib_j)\big) \\
& &\qquad \qquad\Big  \{\frac{R_{ki}}{\pi_k(Q_{ki},\hat\psi_k)} \,
N_{kij}(dt) +\Big (1- \frac{R_{ki}}{\pi_k(Q_{ki},\hat\psi_k)}\Big 
) \hat\rho_{kj} (W_{ki}) \, N_{ki}^x(dt) \Big \}. \nonumber
\end{eqnarray}
The AIPW estimator of $\ib_j$
solves the above equation and is denoted by $\hat\ib_{j,A}$.
%Denote $\hat{\ib}_{A}=(\hat\ib_{1,A}^T,\ldots, \hat\ib_{J,A}^T)^T$.
\comment{\colr The consistency and asymptotic normality of the AIPW estimator $\hat{\ib}_{A}$ are established in Web Appendix A.2. $\hat{\ib}_{A}$ is consistent
if either $r_k(W_{ki})$ or $\rho_{kj}(W_{ki})$ is correctly specified, a double robustness property. 
The distribution of $n^{1/2}\big(\hat{\ib}_{I}-\ib \big)$ converges weakly to $N(0,\iOmega_A)$ if either $r_k(W_{ki})$ or $\rho_{kj}(W_{ki})$ is correctly specified. 
%The covariance matrix of the limiting distribution $\iOmega_A=\iSigma^{-1}\iSigma_A^*\iSigma^{-1}$ is defined in the appendix.
%a consistent estimator for the asymptotic covariance can be given by
%The test based on the asymptotic normal distribution yields a p-value of
When both $r_k(W_{ki})$ and $\rho_{kj}(W_{ki})$ are correctly specified,
Theorem 4  in Web Appendix A.2 shows that $\hat\ib_{j,A}$ is more efficient than $\hat\ib_{j,I}$, $j=1,\dots,J$.
}
%Having obtained the AIPW estimators $\hat\ib_{j,A}$, 
We can similarly estimate the baseline hazard function $\lambda_{0kj}(t)$ by a kernel estimator $\hat \lambda_{0kj}^{A}(t)=\int_0^\tau K_{h}(t-s)\, \hat\Lambda_{0kj}^{A}(ds)$, 
%\begin{equation}
%\label{baseline_AUG} \hat
%\lambda_{0kj}^{A}(t)=\int_0^\tau K_{h}(t-s)\, \hat\Lambda_{0kj}^{A}(ds), \nonumber
%\end{equation}
where
\begin{equation}
\label{dbaseline_AUG}
\hat\Lambda_{0kj}^{A}(t)=\sum_{i=1}^{n_k}\int_0^t \frac{R_{ki}}{\pi_k(Q_{ki},\hat\psi_k)}
\frac{N_{kij}(ds)}{n_k S_{k}^{(0)}(s,\hat\ib_{j,A})}  +\Big (1- \frac{R_{ki}}{\pi_k(Q_{ki},\hat\psi_k)}\Big  ) \,
\frac{\hat\rho_{kj} (W_{ki}) N_{ki}^x(ds) } {n_k S_{k}^{(0)}(s,\hat\ib_{j,A})} \nonumber
\end{equation}
is an estimator of the cumulative baseline function $\Lambda_{0kj}(t)$.

{
\section{Asymptotic properties of IPW and AIPW estimators}
\label{asymptotic}

%\section{Asymptotic properties}
%\label{asymptotics}

%{\colr 
We investigate the asymptotic properties of the IPW estimator $\hat\ib_I=(\hat\ib_{1,I}^T, \ldots, \hat\ib_{J,I}^T)^T$
and the AIPW estimator $\hat{\ib}_{A}=(\hat\ib_{1,A}^T,\ldots, \hat\ib_{J,A}^T)^T$ in this section.
For the theoretical results, we need regularity conditions (A.1)-(A.5), which can be found in the Appendix.
%Conditions (C.1)-(C.4) are standard assumptions for the local linear method under the Cox model with time-varying coefficients (Cai and Sun, 2003). 
%Condition (C.5) is a smoothness condition on the parametric models for $r(\zeta_i, A_i)$   and  $f(A_i |k,T_i,Z_i)$,  comparable to the assumptions used in Gao and Tsiatis (2005) and Lu and Liang (2008).
%}

\subsection{Asymptotic results of inverse probability weighted complete-case estimator}

Let 
\begin{eqnarray}
S^\psi_{ki}&=& \frac{\delta_{ki} (R_{ki}-r_k(W_{ki},\psi_{k0}))}{r_k(W_{ki},\psi_{k0})(1-r_k(W_{ki},\psi_{k0}))}
\frac{\partial r_k(W_{ki},\psi_{k0})}{\partial \psi_k}, \label{score-psi0} \\
I^\psi_k&=& E_k\Big \{\frac{\delta_{ki}}{r_k(W_{ki},\psi_{k0})(1-r_k(W_{ki},\psi_{k0}))}
\frac{\partial r_k(W_{ki},\psi_{k0})}{\partial \psi_k}\Big (\frac{\partial r_k(W_{ki},\psi_{k0})}{\partial \psi_k}\Big )^T \Big \}.\label{info-psi0}
\end{eqnarray}
Then
$S^\psi_{ki}$ and  $I^\psi_k$ be the score vector and
information matrix for $\hat\psi_k$ under (\ref{lik_psi}), with
 $\hat\psi_k-\psi_{k0}=n_k^{-1} \sum_{i=1}^{n_k}  (I^\psi_k)^{-1}S^\psi_{ki}  +o_p(n_{k}^{-1/2})$.

%Let
%\begin{eqnarray}
%{\cal A}_{kij} &=& \int_0^\tau  \big (Z_{ki}(t)-\bar z_k(t, \ib_j)\big) \frac{R_{ki}}{\pi_k(Q_{ki},\psi_{k0})} \,M_{kij}(dt) , \nonumber\\
%{\cal B}_{ki} &=&  \int_0^\tau \big (Z_{ki}(t)\!-\!\bar z_k(t, \ib(u))\big)
%\bigg(1-\frac{R_{ki}}{\pi_k(Q_{ki},\psi_{k0})}\bigg ) E\{M_{kij}(dt) |Q_{ki}\}, \nonumber \\
%{\cal D}_{kj} &=& n_k^{-1}\sum_{i=1}^{n_k} \int_0^\tau \big (Z_{ki}(t)-\bar z_k(t, \ib_j)\big)  \otimes \bigg\{ \frac{-R_{ki}}{(\pi_k(Q_{ki},\psi_{k0}))^2}\frac{\partial\pi_k(Q_{ki},\psi_{k0})}{\partial\psi_k}\,M_{kij}(dt) \bigg\}, \nonumber \\
%{\cal O}_{ki} &=&  {\cal D}_k(I^\psi_k)^{-1}S^\psi_{ki}. \label{ABDO} 
%\end{eqnarray}

The consistency and asymptotic normality of  $\hat\ib_I$ are established in the next two theorems.

\begin{theorem}
\label{consistency_IPW} Under Condition A,
if the model for $r_k(W_{ki})$ is correctly specified, then
$\hat\ib_{j,I}$ $\gop\ib_j$ uniformly in $j=1,\ldots,J$ as $n\to\infty$.
\end{theorem}

\begin{theorem}
\label{normality_IPW} Under Condition A,
if the model for $r_k(W_{ki})$ is correctly specified, then
we have
 %{\colr 
$$n^{1/2}\big(\hat{\ib}_{I}-\ib \big) \god N(0,\iSigma^{-1}\iSigma_I^*\iSigma^{-1}),$$
% }
  as $n\to\infty$,
 where $\iSigma=diag\{\Sigma_j, j=1,\ldots,J\}$ with $\Sigma_j$ given in the condition (A.3),
  $ \iSigma_{I}^* =\sum_{k=1}^K p_k E\big( \ixi_{ki,I}^*\big)^{\otimes 2}$ with
$ \ixi_{ki,I}^* =\big( (\xi_{k1i,I}^*)^T, \ldots,  (\xi_{kJi,I}^*)^T \big)^T$, 
\begin{align*}
\xi_{kji,I}^* = &  \int_0^\tau \big (Z_{ki}(t)-\bar z_k(t,\ib_j) \big)  \frac{R_{ki}}{\pi_k(Q_{ki}) } \,M_{kij}(dt) + D_{kj} (I_k^\psi)^{-1} S_{ki}^\psi ,
\end{align*}
$D_{kj}=E_k {\cal D}_{kj} $, and
\begin{align*}
{\cal D}_{kj} &= n_k^{-1}\sum_{i=1}^{n_k} \int_0^\tau \big (Z_{ki}(t)-\bar z_k(t, \ib_j)\big)  \Big( \frac{-R_{ki}}{(\pi_k(Q_{ki},\psi_{k0}))^2} \Big) \Big(\frac{\partial\pi_k(Q_{ki},\psi_{k0})}{\partial\psi_k} \Big)^T \,M_{kij}(dt) .
\end{align*}
Here  ${\cal D}_{kj} $ is a $p\times q$ matrix. %$\otimes $ is the Cartesian product and
\end{theorem}

\m

%{\colb
Note that the derivative of $U_{I}(j,\ib_j,\hat\psi)$ with respect to $\ib_j$ equals
$$U_{I}^{\prime}(j,\ib_j,\hat\psi)=-\sum_{k=1}^K\sum_{i=1}^{n_k}  \int_0^\tau \frac{R_{ki}}{\pi_k(Q_{ki},\hat\psi_k)}
\tilde J_{k}(t,\ib_j,\hat\psi_k)\, N_{kij}(dt),$$
where $\tilde J_{k}(t,\ib_j,\hat\psi_k)$ is defined at the end of Section \ref{Preliminaries}.
%It can be shown that $\hat\Sigma_{j,I}\equiv -n^{-1}U_{I}^{\prime}(j,\hat\ib_{j,I},\hat\psi) \gop\Sigma_j$ as $n\to\infty$. 
 Let  $\hat\Sigma_{j,I}=-n^{-1}U_{I}^{\prime}(j,\hat\ib_{j,I},\hat\psi )$ and
$\hat{\iSigma}_I=diag\{\hat\Sigma_{j,I}, j=1,\ldots,J\}$.

Let $\hat{\cal D}_{kj}$, $\hat I_k^\psi$, $\hat S_{ki}^\psi$ be the empirical counterparts of 
${\cal D}_{kj}$, $I_k^\psi$, $S_{ki}^\psi$, respectively, obtained by 
replacing expectation with sample average, $\psi_k$ with $\psi_k$, and $M_{kij}(dt)$ with %$\hat M_{kij}(t)$ 
$$\hat M_{kij}(dt) = N_{kij}(dt) -  Y_{ki}(t)  \exp(\hat\ib_{j,I}^{T} Z_{ki}(t) )\,d \hat\Lambda_{0kj}^I(t).$$
%}
 
%{\colr
 Let
$ \tilde \iSigma_{I}^* = n^{-1}\sum_{k=1}^K  \sum_{i=1}^{n_k} \big ( \tilde \ixi_{ki,I}^* \big)^{\otimes 2}$,
 $ \tilde \ixi_{ki,I}^* =\big( (\tilde \xi_{k1i,I}^*)^T, \ldots,  (\tilde \xi_{kJi,I}^*)^T \big)^T$ and

\begin{align*}
 \tilde \xi_{kji,I}^* = &  \int_0^\tau \big[Z_{ki}(t)- \tilde Z_k(t,\hat\ib_{j,I},\hat\psi_k)\big ] 
\frac{R_{ki}}{(\pi_k(Q_{ki},\hat\psi_k))^2} \hat M_{kij}(dt) + \hat{\cal D}_{kj} (\hat I_k^\psi)^{-1} \hat S_{ki}^\psi .
\end{align*}

By the consistency of $\hat{\ib}_{I}$ and  $\hat\psi$,  $\iSigma$ and $ \iSigma_{I}^*$
can be consistently estimated by
$\hat{\iSigma}_I$ and $ \tilde \iSigma_{I}^* $, respectively.
%}

Under Theorem \ref{normality_IPW},
we have
 $n^{1/2}\big(\hat\ib_{j,I}-\ib_j\big)$
 $\god N(0,\Sigma_j^{-1}\Sigma_{j,I}^*\Sigma_j^{-1})$,
  where   $\Sigma_{j,I}^*=\sum_{k=1}^K p_k E_k\big( \xi_{kji,I}^*\big)^{\otimes 2}$ for $j=1,\ldots,J$ as $n\to\infty$.
Let $ \tilde \Sigma_{j,I}^* = n^{-1}\sum_{k=1}^K  \sum_{i=1}^{n_k} \big ( \tilde \xi_{kji,I}^* \big)^{\otimes 2}$.
The asymptotic variance of $n^{1/2}(\hat\ib_{j,I}-\ib_j)$ can be
consistently estimated by $(\hat\Sigma_{j,I})^{-1}$ $\tilde\Sigma_{j,I}^*(\hat\Sigma_{j,I})^{-1}$.
The IPW estimators $\hat\ib_{j,I}$, $j=1,\ldots,J$, are not asymptotically independent.

\subsection{Asymptotic results of augmented inverse probability weighted complete-case estimator}

%%%
\comment{
Let $S^\psi_{ki}$ and  $I^\psi_k$ be the score vector and
information matrix for $\hat\psi_k$ under (\ref{lik_psi}). Then
\begin{eqnarray}
S^\psi_{ki}&=& \frac{\delta_{ki} (R_{ki}-r_k(W_{ki},\psi_{k0}))}{r_k(W_{ki},\psi_{k0})(1-r_k(W_{ki},\psi_{k0}))}
\frac{\partial r_k(W_{ki},\psi_{k0})}{\partial \psi_k}, \label{score-psi0} \\
I^\psi_k&=& E\Big \{\frac{\delta_{ki}}{r_k(W_{ki},\psi_{k0})(1-r_k(W_{ki},\psi_{k0}))}
\frac{\partial r_k(W_{ki},\psi_{k0})}{\partial \psi_k}\Big (\frac{\partial r_k(W_{ki},\psi_{k0})}{\partial \psi_k}\Big )^T \Big \},\label{info-psi0}
\end{eqnarray}
and $\hat\psi_k-\psi_{k0}=n_k^{-1} \sum_{i=1}^{n_k}  (I^\psi_k)^{-1}S^\psi_{ki}  +o_p(n_{k}^{-1/2})$.
}
%%%

We  introduce the following notation:
\begin{eqnarray}
{\cal A}_{kij} &=& \int_0^\tau \big (Z_{ki}(t)-\bar z_k(t, \ib_j)\big) \frac{R_{ki}}{\pi_k(Q_{ki},\psi_{k0})} \, M_{kij}(dt), \nonumber\\
{\cal B}_{kij} &=& \int_0^\tau  \big (Z_{ki}(t)- \bar z_k(t, \ib_j)\big)
\Big (1-\frac{R_{ki}}{\pi_k(Q_{ki},\psi_{k0})}\Big  )E\{M_{kij}(dt)|Q_{ki}\}, \nonumber \\
{\cal D}_{kj} &=& n_k^{-1}\sum_{i=1}^{n_k}  \int_0^\tau \big (Z_{ki}(t)-\bar z_k(t, \ib_j)\big) \otimes \Big \{ \frac{-R_{ki}}{(\pi_k(Q_{ki},\psi_{k0}))^2}\frac{\partial\pi_k(Q_{ki},\psi_{k0})}{\partial\psi_k}
\,M_{kij}(dt)\Big \},  \nonumber \\
{\cal O}_{kij} &=&  {\cal D}_{kj}(I^\psi_k)^{-1}S^\psi_{ki}. \label{ABDO} 
\end{eqnarray}

The next theorem shows that the AIPW estimator $\hat\ib_{j,A}$ is consistent
if either $r_k(w,\psi_k)$ or $g_k(a|t,v,$ $z,\theta_k)$ is correctly specified,
a double robustness property.

\begin{theorem}
\label{consistency_DR} Assuming Condition A,
$\hat\ib_{j,A}\gop \ib_j$ uniformly in $j=1,\ldots,J$
as $n\to\infty$.
This consistency holds if either $r_k(w,\psi_k)$ or $g_{kj}(a|t,z,\theta_k)$
is correctly specified.
\end{theorem}

%%%
\comment{
Let
\begin{align*}
 \tilde \xi_{kji,A}^* = &  \int_0^\tau \big(Z_{ki}(t)-\bar Z_k(t,\hat\ib_{j,A})\big) \\
& \Big  [\frac{R_{ki}}{\pi_k(Q_{ki},\hat\psi_k)} \,N_{kij}(dt) +\Big (1- \frac{R_{ki}}{\pi_k(Q_{ki},\hat\psi_k)}\Big 
) \hat\rho_{kj}^{I}(W_{ki}) \, N_{ki}^x(dt) \Big  ],
\end{align*}
$ \tilde \ixi_{ki,A}^* =\big( (\tilde \xi_{k1i,A}^*)^T, \ldots,  (\tilde \xi_{kJi,A}^*)^T \big)^T$ and
$ \tilde \iSigma_{A}^* = n^{-1}\sum_{k=1}^K  \sum_{i=1}^{n_k} \big ( \tilde \ixi_{ki,A}^* \big)^{\otimes 2}.$
\m

By the consistency of $\hat{\ib}_{A}$, $\hat\psi$, and $\hat\rho^{I}(w)$, it follows that
$ \tilde \iSigma_{A}^* \gop  \iSigma_{A}^* =\sum_{k=1}^K p_k \Sigma_{k}^*$,
where
$\Sigma_k^* =E\big( \ixi_{ki,A}^*\big)^{\otimes 2}$,
$ \ixi_{ki,A}^* =\big( (\xi_{k1i,A}^*)^T, \ldots,  (\xi_{kJi,A}^*)^T \big)^T$, and
\begin{align*}
\xi_{kji,A}^* = &  \int_0^\tau \big (Z_{ki}(t)-\bar z_k(t,\ib_j) \big)   \Big  [\frac{R_{ki}}{\pi_k(Q_{ki}) } \,N_{kij}(dt) +\Big (1- \frac{R_{ki}}{\pi_k(Q_{ki}) }\Big 
) \rho_{kj}(W_{ki}) \, N_{ki}^x(dt) \Big  ].
\end{align*}

}
%%%

\m

Following the proofs of Theorem \ref{normality_IPW} and \ref{consistency_DR},
it is easy to show that $n^{1/2}\big(\hat\ib_{j,A}-\ib_j\big)$ is asymptotically normal for $j=1,\ldots,J$ 
if either $r_k(w,\psi_k)$ or $g_{kj}(a|t,z,\theta_k)$ is correctly specified.
When both $r_k(w,\psi_k)$ and $g_{kj}(a|t,z,\theta_k)$ are correctly specified,
Theorem \ref{normality_DR} below shows that
$\hat\ib_{j,A}$ is more efficient than $\hat\ib_{j,I}$.

\begin{theorem}
\label{normality_DR} Assuming Condition A,
if both $r_k(w,\psi_k)$ and $g_{kj}(a|t,z,\theta_k)$ are correctly specified for $j=1,\ldots,J$ and for $k=1,\ldots,K$,
we have
 $$n^{1/2}\big(\hat{\ib}_{A}-\ib \big) \god N(0,\iSigma^{-1}\iSigma_A^*\iSigma^{-1}),$$  as $n\to\infty$,
 where $\iSigma=diag\{\Sigma_j, j=1,\ldots,J\}$
with $\Sigma_j$ given in the condition (A.3),
  $ \iSigma_{A}^* =\sum_{k=1}^K p_k E\big( \ixi_{ki,A}^*\big)^{\otimes 2}$ with
% $ \tilde \iSigma_{A}^* \gop  \iSigma_{A}^* =\sum_{k=1}^K p_k \Sigma_{k}^*$,
%$\iOmega_{k,A}^* =E\big( \ixi_{ki,A}^*\big)^{\otimes 2}$,
$ \ixi_{ki,A}^* =\big( (\xi_{k1i,A}^*)^T, \ldots,  (\xi_{kJi,A}^*)^T \big)^T$, and
%{\colb
\begin{align*}
\xi_{kji,A}^* & =   \int_0^\tau \big (Z_{ki}(t)-\bar z_k(t,\ib_j) \big)   \Big  [\frac{R_{ki}}{\pi_k(Q_{ki}) } \,M_{kij}(dt) +\Big (1- \frac{R_{ki}}{\pi_k(Q_{ki}) }\Big )E\{M_{kij}(dt)|Q_{ki}\} \Big  ].
%
%&  =   \int_0^\tau \big (Z_{ki}(t)-\bar z_k(t,\ib_j) \big)   \Big  [\frac{R_{ki}}{\pi_k(Q_{ki}) } \,N_{kij}(dt) +\Big (1- \frac{R_{ki}}{\pi_k(Q_{ki}) }\Big ) \rho_{kj}(W_{ki}) \, N_{ki}^x(dt) \Big  ]\\
%&  \quad -  \int_0^\tau \big (Z_{ki}(t)-\bar z_k(t,\ib_j) \big)  Y_{ki}(s)  \exp(\ib_j^{T} Z_{ki}(s) )\,d \Lambda_{0kj}(s)
\end{align*}
%}
\end{theorem}

\b

Note that $M_{kij}(t)=\int_0^t [N_{kij}(ds)-Y_{ki}(s)\lambda_{kj}(s|Z_{ki}(s))\,ds]$
and $\lambda_{kj}(t|Z_{ki}(t))= \lambda_{0kj}(t) \exp(\ib_j^{T} Z_{ki}(t) )$.
We have 
%{\colr
\begin{align*}
\xi_{kji,A}^* 
%& =   \int_0^\tau \big (Z_{ki}(t)-\bar z_k(t,\ib_j) \big)   \Big  [\frac{R_{ki}}{\pi_k(Q_{ki}) } \,M_{kij}(dt) +\Big (1- \frac{R_{ki}}{\pi_k(Q_{ki}) }\Big )E\{M_{kij}(dt)|Q_{ki}\} \Big  ].\\
%
  =   \int_0^\tau \big (Z_{ki}(t)-\bar z_k(t,\ib_j) \big) &  \Big  [\frac{R_{ki}}{\pi_k(Q_{ki}) } \,N_{kij}(dt) +\Big (1- \frac{R_{ki}}{\pi_k(Q_{ki}) }\Big 
) \rho_{kj}(W_{ki}) \, N_{ki}^x(dt) \\
&  -    Y_{ki}(s)  \exp(\ib_j^{T} Z_{ki}(s) )\,d \Lambda_{0kj}(s) \Big  ].
\end{align*}
%}

\comment{
Let $\hat\rho_{kj}^{A}(w)$ be defined similarly to $\hat\rho_{kj}^{I}(w)$
given in (\ref{rho-ipw}) using the AIPW estimator  $\hat\lambda_{kj}^{A}(t|z)=
\hat\lambda_{0kj}^{A}(t)\exp\{(\hat\ib_{j,A})^Tz\}$.
Let $\hat\rho_j^{A}(w)=(\hat\rho_{1j}^{A}(w), \ldots, \hat\rho_{Kj}^{A}(w))$,
$\hat\Sigma_{j,A}=-n^{-1}U_{A}^{\prime}(j,\hat\ib_{j,A},\hat\psi,\hat\rho_j^{A}(\cdot))$ and
$\hat{\iSigma}_A=diag\{\hat\Sigma_{j,A}, j=1,\ldots,J\}$.
}

Let $\hat\Sigma_{j,A}=-n^{-1}U_{A}^{\prime}(j,\hat\ib_{j,A},\hat\psi,\hat\rho(\cdot))$ and
$\hat{\iSigma}_A=diag\{\hat\Sigma_{j,A}, j=1,\ldots,J\}$.
Let
$ \tilde \iSigma_{A}^* = n^{-1}\sum_{k=1}^K  \sum_{i=1}^{n_k} \big ( \tilde \ixi_{ki,A}^* \big)^{\otimes 2}$,
 $ \tilde \ixi_{ki,A}^* =\big( (\tilde \xi_{k1i,A}^*)^T, \ldots,  (\tilde \xi_{kJi,A}^*)^T \big)^T$ and
% {\colr
\begin{align*}
 \tilde \xi_{kji,A}^* =   \int_0^\tau & \big(Z_{ki}(t)-\bar Z_k(t,\hat\ib_{j,A})\big) \\
& \quad \Big  [\frac{R_{ki}}{\pi_k(Q_{ki},\hat\psi_k)} \,N_{kij}(dt) +\Big (1- \frac{R_{ki}}{\pi_k(Q_{ki},\hat\psi_k)}\Big 
) \hat\rho_{kj} (W_{ki}) \, N_{ki}^x(dt) \\
 & \quad -  Y_{ki}(s)  \exp(\hat\ib_{j,A}^{T} Z_{ki}(s) )\,d \hat\Lambda_{0kj}^A(s) \Big  ].
\end{align*}
%}

\m

By the consistency of $\hat{\ib}_{A}$, $\hat\psi$, and $\hat\rho(\cdot)$,  $\iSigma$ and $ \iSigma_{A}^*$
can be consistently estimated by
$\hat{\iSigma}_A$ and $ \tilde \iSigma_{A}^* $, respectively.

Under Theorem \ref{normality_DR},
we have
 $n^{1/2}\big(\hat\ib_{j,A}-\ib_j\big)$
 $\god N(0,\Sigma_j^{-1}\Sigma_{j,A}^*\Sigma_j^{-1})$ for $j=1,\ldots,J$ as $n\to\infty$,
 where $\Sigma_{j,A}^*=\sum_{k=1}^K p_k E\big( \xi_{kji,A}^*\big)^{\otimes 2}$.
Let $ \tilde \Sigma_{j,A}^* = n^{-1}\sum_{k=1}^K  \sum_{i=1}^{n_k} \big ( \tilde \xi_{kji,A}^* \big)^{\otimes 2}$.
The asymptotic variance of $n^{1/2}(\hat\ib_{j,A}-\ib_j)$ can be
consistently estimated by $(\hat\Sigma_{j,A})^{-1}$ $\tilde\Sigma_{j,A}^*(\hat\Sigma_{j,A})^{-1}$.
The AIPW estimators $\hat\ib_{j,A}$, $j=1,\ldots,J$, are not asymptotically independent.

%{\colr (LATER)} 
The estimator $\hat\ib_{j,A}$ is more efficient than
$\hat\ib_{j,I}$ in the sense that
%{\colb 
\begin{align}
\label{efficient}
 Cov\{n^{1/2} \big(\hat\ib_{j,I}-\ib_j \big) \}
&=Cov\{n^{1/2} \big(\hat\ib_{j,A}-\ib_j \big)\}\nonumber\\
&  + \Sigma_j^{-1}\Big (\sum_{k=1}^K ({n_k}/{n}) Cov\{{\cal O}_{k1}-{\cal B}_{k1}\}\Big  )\Sigma_j^{-1}
+o_p(1).
\end{align}
%}
Equation (\ref{efficient}) shows that the asymptotic covariance
$ Cov\{n^{1/2} \big(\hat\ib_{j,A}-\ib_j \big) \}$
is smaller than $ Cov\{n^{1/2} \big(\hat\ib_{j,I}-\ib_j \big) \}$. 
%The demonstrated efficiency gain for $\hat\ib_{j,A}$
%is reasonable since the estimation procedures for both $\hat\ib_{j,A}$ and
%$\hat\ib_{j,I}$ are based on the local partial likelihood with $h$ as its bandwidth.
}

\section{Statistical inferences for vaccine efficacy}
\label{inference}

%{\colr 
Under the stratified cause-specific Cox model (\ref{causeM1}), the strain-specific vaccine efficacy $VE_j=1-\exp(\alpha_j)$, where $\alpha_j$ is the first component of the covariate coefficient vector $\ib_j$, representing the coefficient for vaccination status.
Confidence intervals and hypothesis testing procedures for $\{VE_j, j=1,\ldots,J\}$ are constructed on the basis of estimators $\ib_{I}$ and $\ib_{A}$ obtained in Section \ref{estimation}. For simplicity, we omit the subscripts $I$ and $A$ and generally use $\widehat\ib=(\widehat\ib_1,\dots,\widehat\ib_J)^T$ for the estimator of $\ib=(\ib_1,\dots,\ib_J)^T$, $\iOmega$ for the covariance matrix of the limiting distribution of $n^{1/2}\big(\hat{\ib}-\ib \big)$, and $\widehat\iap=(\widehat\alpha_1,\dots,\widehat\alpha_J)^T$ for the estimator of $\iap=(\alpha_1,\dots,\alpha_J)^T$.
%}

\subsection{Confidence intervals}
\label{confint}

%{\colr 
By the asymptotic results in Section \ref{asymptotic}, $n^{1/2} \big(\widehat\iap -\iap \big) \god N(0, \iOmega_{\iap}),$  as $n\to\infty$,
where $\iOmega_{\iap}$ is the asymptotic covariance matrix consists of elements in corresponding positions of $\iOmega$.
%Further, the limiting covariance matrix can be consistently estimated according to the consistency of estimated parameters. 
Let $\widehat\iOmega_{\iap}$ be a consistent estimator of $\iOmega_{\iap}$ and $\widehat\iOmega_{\iap,ij}$ be the $(i,j)$th entry of $\widehat\iOmega_{\iap}$.
A large sample $100(1-\alpha)\%$ confidence interval for $\alpha_j$ is given by
$\hat\alpha_j \pm z_{\alpha/2} \widehat\sigma_{j}, j=1,\ldots,J,$
where $z_{{\alpha}/2}$ is the upper $\alpha/2$th percentile of the standard normal distribution and $\widehat \sigma_j=(\widehat\iOmega_{\iap,jj}/n)^{1/2}$ is an estimate of $\sigma_j$, the standard error of $\widehat\alpha_j$. 
%The confidence intervals for the other coefficients can be constructed similarly.

The strain-specific vaccine efficacy $VE_j=1-\exp( \alpha_j)$ can be
estimated by $\widehat  {VE}_j=1-\exp(\hat\alpha_j)$.
By the asymptotic property of $\alpha_j$ and the delta method, we have $n^{1/2}(\widehat  {VE}_j- VE_j) \god N(0,\sigma_j^2\exp(2 \alpha_j))$ for $j=1\ldots,J$.
An approximate $100(1-\alpha)\%$ confidence interval for $ VE_j$ is then given by 
$\widehat  {VE}_j \pm z_{\alpha/2} \widehat\sigma_{j} \exp(\hat\alpha_j), j=1\ldots,J.$
Using the transformation $\log((1-\hat{VE})/(1-VE))=\hat\alpha_j-\alpha_j$, we can construct an alternative large-sample approximation of the $100(1-\alpha)\%$ confidence interval for $ VE_j$:
$[1-\exp(\hat\alpha_j + z_{\alpha/2}\widehat\sigma_{j}), 1-\exp(\hat\alpha_j - z_{\alpha/2}\widehat\sigma_{j})], j=1\ldots,J.$
Our numerical studies show that the latter one has better coverage probability.
%Our numerical studies show that the confidence interval for $VE_j$ obtained  using the transformation $\log((1-\hat{VE})/(1-VE))$ has better coverage probability.

%Point estimate and 95\% CI for (1 ? VE(1)) / (1 ? VE(2)), which measures how much greater the vaccine protection is against a V=1 virus than against a V=2 virus.

To measure how much greater the level of VE is against a strain $V=j$ virus than against a strain $V=i$ virus, we define 
$$VD(i,j)=\frac{1-VE_i}{1-VE_j}=\exp(\alpha_i-\alpha_j),$$
%{\colb $$VD_{21}=\frac{1-VE_2}{1-VE_1}=\exp(\alpha_2-\alpha_1)?$$}
which can be estimated by $\widehat{VD}(i,j)=\exp(\widehat\alpha_i-\widehat\alpha_j)$.
A larger $VD(i,j)$ value indicates that the vaccine provides greater protection against a strain type $j$ virus than against a strain type $i$ virus.
The asymptotic variance of $\widehat\alpha_i-\widehat\alpha_j$ can be estimated by $\widehat{Var}(\ia_{i}-\ia_{j})=n^{-1}(\widehat\iOmega_{\iap,ii}+\widehat\iOmega_{\iap,jj}-2\widehat\iOmega_{\iap,ij})$.
%Let $\hat\sigma_{ij}$ estimated standard deviation of $\widehat\alpha_2-\widehat\alpha_1$.
%by Theorem \ref{normality_DR} and the delta method, 
%$$\frac{\widehat {VD}_{12}- VD_{12}}{\sqrt{Var(\widehat\alpha_1^A-\widehat\alpha_2^A)}} \god N\big(0, \exp(2 (\alpha_1-\alpha_2)) \big).$$ 
A large sample $100(1-\alpha)\%$ confidence interval for $VD(i,j)$ using the logarithm transformation is
$$\left[\widehat{VD}(i,j)\exp\left(-z_{\alpha/2}\sqrt{\widehat{Var}(\ia_{i}-\ia_{j})}\right), \widehat{VD}(i,j)\exp\left(z_{\alpha/2}\sqrt{\widehat{Var}(\ia_{i}-\ia_{j})}\right)\right].$$
  
%$$ \widehat{VD} \pm   z_{\alpha/2} \sqrt{ \widehat{Var}(\widehat\alpha_1^A-\widehat\alpha_2^A)} \widehat{VD}.$$

%\vskip 0.5in
%Our numerical studies show that the confidence interval for $VE_j$ obtained  using the transformation $\log(\hat{VD}/VD)$ has better coverage probability. Similarly,
%the confidence interval for $VD$ obtained  using the transformation $\log((1-\hat{VE})/(1-VE))$ has better coverage probability.
%}

\subsection{Testing strain-specific vaccine efficacy}
\label{testing}

We propose test procedures to evaluate various hypotheses concerning strain-specific VE. The tests assess if the vaccine provides at least a certain specified
level of efficacy against some strains and whether vaccine efficacy varies across strains.
The hypothesis tests concerning $VE_j$ are constructed based on the estimator of $\alpha_{j}$.
%The hypothesis tests concerning $VE_j$ are constructed based on the first component $\widehat\alpha_{j,A}$ of the AIPW estimator $\widehat\ib_{j,A}$.

(1) 
First, we consider testing $VE_j \le VE_0$ for $j=1,\ldots,J$, where $VE_0$ is a fixed constant such 
as 0.30 or 0. Let $c_0=\log(1-VE_0)$.
%{\colb Fei, in the Code, please set the constant $c_0$ for flexibility. We already have $c_0=0$.
%We might also want to try some cases with $c_0=\log(1-VE_0)=\log(1-0.3)=-0.3566749$.}
We develop the tests of the null hypothesis (A) that VE is at most $VE_0$ against all strains 
$H_{A0}: VE_j \le VE_0 \mbox{ for all } j=1,\cdots, J$.
This is equivalent to testing
\[
  \begin{array}{cl}
    H_{A0}:& \alpha_{j} \ge c_0, \mbox{ for all } j=1,\cdots, J \\
  \end{array}
\]
versus one of the following alternative hypotheses
\[
  \begin{array}{cl}
    H_{A1}:& \alpha_{j}\leq c_0 \mbox{ with strict inequality for
    some } j, \\
    H_{A2}:& \alpha_{j} \neq c_0 \mbox{ for some } j.
  \end{array}
\]
Thus, $H_{A0}$ implies that VE against any strain is no more than $VE_0$, say, 
30\%. The alternative $H_{A1}$ indicates that the VE is higher than $VE_0$  for at least some of the viral strains, 
while $H_{A2}$ states that VE differs from $VE_0$ for some of the viral strains.

The following test statistics are proposed for detecting
departures from $H_{A0}$ in the directions of $H_{A1}$ and $H_{A2}$, respectively:
\begin{eqnarray}
        U_1=  \inf_{1 \leq j \leq J} \dfrac{\ia_j -c_0 }{\isg_j}, %\label{U1}  \\
                U_2 = \sum_{j=1}^J  \Big(\dfrac{\ia_j-c_0 }{\isg_j} \Big)^2.  \nonumber %\label{U2a}                   
%        U_2 &=&  {\colr  n \sup_{1 \leq j \leq J} \Big(\dfrac{\ia_j-c_0 }{\isg_j} \Big)^2 } \label{U2} \\
%        U_{2a} &= & {\colr n ( \widehat\alpha_{A} -c_0 \mathbf{1} )^T\hat\iOmega^{-1} (\widehat\alpha_{A}-c_0 \mathbf{1} ) } \; \text{Or use?} = n \sum_{j=1}^J  \Big(\dfrac{\ia_j-c_0 }{\isg_j} \Big)^2,  \label{U2a}   
\end{eqnarray}
The test statistic $U_1$ can be used to detect the departure $H_{A1}$ from $H_{A0}$
and $U_2$ can be used to detect the general departure $H_{A2}$ from $H_{A0}$.
Under $H_{A0}$, the test statistic $U_1$ has the asymptotic
distribution of $\inf_{1 \leq j \leq J} Z_j/{\sigma_j} $ and $U_2$ has the asymptotic
distribution of $\sum_{j=1}^J  Z_j^2/{\sigma_j^2} $, where the vector $(Z_1,\ldots,Z_J)$ follows a multivariate normal distribution with mean vector $\boldsymbol 0 = (0,\dots, 0)$ and  covariance matrix $\iOmega_{\alpha})$. 
Let $(\widehat Z_1,\ldots,\widehat Z_J) \sim N(0, \widehat\iOmega_\alpha)$.
%{\colr 
Let $U_{1, \alpha}^*$ be the $\alpha$th percentile of $U_1^*=\inf_{1 \leq j \leq J} \widehat Z_j/ {\isg_j}$, 
and $U_{2, \alpha}^*$ the upper $\alpha$th percentile of $U_2^*=\sum_{j=1}^J  \widehat Z_j^2/ {\isg_j}^2 $, respectively.
%} 
If $U_1 < U_{1, \alpha}^*$, the test based on the test statistic $U_1$ rejects $H_{A0}$ in favor of the alternative
$H_{A1}$ at the $\alpha$ level of significance. If $U_2 > U_{2, \alpha}^*$,
the test based on the test statistic $U_2$ rejects $H_{A0}$ in favor of the alternative
$H_{A2}$.

(2)
\comment{\colr
We also present multiple comparison pocedures designed to test VE against each strain $j$, $ j=1,\cdots, J$. We test the following hypotheses:
$$H_{Aj0}:  VE_j \le VE_0 \; (\alpha_{j} \ge c_0)$$%, \quad H_{Aj1}: VE_j \ge VE_0 \; (\alpha_{j} \le c_0)$$
versus one of the following alternative hypotheses
\[
  \begin{array}{cl}
    H_{Aj1}:& VE_j > VE_0\;  (\alpha_{j}< c_0), \\
    H_{Aj2}:& VE_j \neq VE_0\;  (\alpha_{j} \neq c_0).
  \end{array}
\]
The test statistics $U_{1j}=(\ia_j -c_0)/{\isg_j}$ and $ U_{2j}=(\ia_j-c_0)^2/{\isg_j^2}$ are used for detecting departures from $H_{Aj0}$ in the directions of $H_{Aj1}$ and $H_{Aj2}$, respectively. 
We can approximate distributions of $U_{1j}$ and $ U_{2j}$ using $(\widehat Z_1,\ldots,\widehat Z_J)$ similar to above, and can further compute unjusted p-values.
A step-down \v{S}id\'{a}k-like method can be leveraged to control for the family-wise error rate (Holland and Copenhaver, 1987).
The corresponding adjusted p-values are given by $p_{(j)}^{adj}=\max_{i\le j}\{1-(1-p_{(i)})^{J+1-i}\}$, where $\{p_{(i)}; i=1,\dots, J\}$ are ordered unadjusted $p$-values from smallest to largest.
}
{
%For the Moderna COVID vaccine trial, we are also interested in testing $VE_j=1-\exp(\alpha_j)\le VE_0$ for each $j=1,\ldots,J$, where $VE_0=0.30$ or 0. Let $c_0=\log(1-VE_0)$.
To test VE against each strain $j$, $ j=1,\dots, J$, we test the following hypotheses:
$$H_{Aj0}:  VE_j \le VE_0 \; (\alpha_{j} \ge c_0)$$%, \quad H_{Aj1}: VE_j \ge VE_0 \; (\alpha_{j} \le c_0)$$
versus one of the following alternative hypotheses
\[
  \begin{array}{cl}
    H_{Aj1}:& VE_j > VE_0\;  (\alpha_{j}< c_0), \\
    H_{Aj2}:& VE_j \neq VE_0\;  (\alpha_{j} \neq c_0).
  \end{array}
\]
The following test statistics are used for detecting departures from $H_{Aj0}$ in the directions of $H_{Aj1}$ and $H_{Aj2}$, respectively:
\begin{eqnarray}
        U_{1j} =  \dfrac{\ia_j -c_0 }{\isg_j}, %\label{Uj1} \\
        U_{2j} =  \left(\dfrac{\ia_j-c_0 }{\isg_j} \right)^2. \nonumber %\label{Uj2} 
\end{eqnarray}
The critical values are obtained similar to above but only for one $j$ at a time.
%{\colr
In clinical trials that require the simultaneous test of $J$ null hypotheses $\{H_{Aj0},  j=1,\dots, J\}$, a common approach is to apply Bonferroni adjustment of the level of significance. 
While the Bonferroni procedure is simple to implement, it tends to be quite conservative for control of the familywise error rate (FWER), especially when the number of tests is large.
Since the test statistics follow multivariate Gaussian distribution asymptotically, we can apply a less conservative step-down \v{S}id\'{a}k-like procedure here (Holland and Copenhaver, 1987). \nocite{HollandCopenhaver1987}
The corresponding adjusted p-values are given by $p_{(j)}^{adj}=\max_{i\le j}\{1-(1-p_{(i)})^{J+1-i}\}$, where $\{p_{(i)}; i=1,\dots, J\}$ are ordered unadjusted $p$-values from smallest to largest.

(3) Next, we develop tests for whether strain-specific VE depends on strain type, so-called ``sieve effect" tests.
These tests evaluate the null hypothesis (B) 
\[H_{B0}: VE_1=VE_2= \cdots =VE_J\] 
versus the following alternative hypotheses:
\[
  \begin{array}{cl}
    H_{B1}:& VE_1 \geq \cdots \geq VE_j \geq \cdots \geq VE_J
    \mbox{ with strict inequality for some } 1\leq j \leq J, \\
    H_{B2}:& VE_i \neq VE_j  \mbox{ for at least one pair of
    } i \mbox{ and } j, \quad 1\leq i < j\leq J.
  \end{array}
\]
$H_{B0}$ implies that strain-specific VE does not vary with strain type.
The ordered alternative $H_{B1}$ states that VE
decreases with strain type. Under the proportional hazards model
(\ref{causeM1}) the hypotheses (B) can be rewritten as
\[
\begin{array}{cl}
    H_{B0}:& \alpha_{1}=\alpha_{2}=\cdots=\alpha_J
\end{array}
\]
against the following alternative hypotheses
\[
  \begin{array}{cl}
    H_{B1}:& \alpha_{1} \leq \cdots \leq \alpha_{j}\leq \cdots \leq
    \alpha_J  \mbox{ with at least one strict inequality},
    \\
    H_{B2}:& \alpha_{i} \neq \alpha_j  \mbox{ for at least one pair of
    } i \mbox{ and } j, \quad 1\leq i < j\leq J.
  \end{array}
\]
\comment{
We develop two test statistics $T_1$ and $T_2$ for detecting departures from
$H_{B0}$ in the direction of $H_{B1}$ and $H_{B2}$, respectively:
\begin{eqnarray}
   T_{1} &=&  \inf_{2 \le j \le J} \frac{\ia_{j}-\ia_{j-1}}{\sqrt{\widehat{Var}(\ia_{j}-\ia_{j-1})}}, \\
   T_{2} &=& \sum_{j=2}^J \frac{(\ia_{j}-\ia_{j-1})^2 }{\widehat{Var}(\ia_{j}-\ia_{j-1})},
\end{eqnarray}
where $\widehat{Var}(\ia_{j}-\ia_{j-1})$ is the estimate of ${Var}(\ia_{j}-\ia_{j-1})={Var}(\ia_{j})-2{Cov}(\ia_{j-1},\ia_{j}) + {Var}(\ia_{j-1})$ , 
which can be obtained from $\hat\iOmega_{\alpha}$.
}

The following test statistic $T_1$ is suggested for detecting the monotone departure $H_{B1}$ from $H_{B0}$: 
\begin{eqnarray}
   T_{1} &=&  \inf_{2 \le j \le J} \frac{\ia_{j}-\ia_{j-1}}{\sqrt{\widehat{Var}(\ia_{j}-\ia_{j-1})}}, \nonumber
\end{eqnarray}
where $\widehat{Var}(\ia_{j}-\ia_{j-1})$ is the estimate of ${Var}(\ia_{j}-\ia_{j-1})={Var}(\ia_{j})-2{Cov}(\ia_{j-1},\ia_{j}) + {Var}(\ia_{j-1})$ , 
which can be obtained from $\hat\iOmega_{\alpha}$.
Under $H_{B0}$, asymptotically, we can approximate $T_1$ by  %normal with mean zero and variance of one. 
$T_1^*=\inf_{2 \le j \le J} ({\hat Z_{j}-\hat Z_{j-1}})/{\sqrt{\widehat{Var}(\ia_{j}-\ia_{j-1})}}.$
The $H_{B0}$ is rejected in favor of $H_{B1}$ at significance level $\alpha$
if $T_1 > T_{1,\alpha}^*$, where $T_{1,\alpha}^*$ is the upper $\alpha$th percentile of $T_1^*$. 
To detect general alternative $H_{B2}$ from $H_{B0}$, we consider the test statistic
$$
     T_{2} = \sum_{j=2}^J \frac{(\ia_{j}-\ia_{j-1})^2} {\widehat{Var}(\ia_{j}-\ia_{j-1})}. 
$$
The asymptotic distribution of test statistic $T_2$ under $H_{B0}$ can be approximated by the distribution of
$T_2^*=\sum_{j=2}^J {(\hat Z_{j}-\hat Z_{j-1})^2} /{\widehat{Var}(\ia_{j}-\ia_{j-1})}.$
The $H_{B0}$ is rejected in favor of $H_{B2}$ if
$T_2 >  T_{2,\alpha}^*$, where $ T_{2,\alpha}^*$ is the upper $\alpha$th percentile of $T_2^*$. 

\section{Simulation study}
\label{simulation}

We conduct a simple simulation study to examine the performance of the proposed methods.
We consider a $p=2$ dimensional covariate $Z=(Z_1,Z_2)$, where $Z_1$ is a Bernoulli random variable 
with probability of success 0.5 that represents the treatment group indicator,
and $Z_2$ is a uniformly distributed random variable on $(0,1)$.
We consider the following cause-specific proportional hazards model for $J=2$ causes and $K=3$ strata:
\begin{equation}
\label{simu-model1}
 \lambda_{kj}(t|z)=t^{\theta_{kj}}\exp(\alpha_{j}z_{1}+\gamma_{j}z_{2}), \quad j=1,2, \quad k=1,2,3, 
\end{equation}
where $\theta_{kj}$, $\alpha_j$ and $\gamma_j$ are the parameters to be specified.
All failure times greater than $\tau=1$ are right-censored at $\tau$.
In addition, random censoring times are generated from an exponential distribution,
independent of $(T,V)$, with parameter adjusted so approximately 40\% of the observations are censored.
The sizes and powers of the tests at the nominal 0.05 level are estimated from 1000 independent samples. 

We consider a single auxiliary covariate $A$ that follows a Bernoulli distribution with success probability of 
$0.5$.  
For the cause $V=j$, we also generate a single auxiliary mark variable $A$ 
that follows a uniform distribution on $(2a(j-1),1+0.5aj)$. 
%
%{\colr Let  $A$ follow a uniform distribution on $(j-a,j+a)$, where we can take, e.g., $a=0.6, 0.7$?}
%
We examine the performance of the estimators under three different levels of association between $A$ and the failure 
cause $V$, 
by considering the settings $a=0,0.2,$ and $0.5$, which result in approximate Kendall's tau values of 
$0, 0.3,$ and $0.6$, respectively.  
These three auxiliary association level settings are denoted by (Aux0), (Aux1), and (Aux2), respectively.
Note that  $A$ is independent of $V$ for the setting (Aux0), and the association between $A$ and $V$ 
increases from (Aux1) to (Aux2). 

The cause $V$ is missing at random (MAR). $r_k(W)$, the conditional probability that the cause 
is not missing when $\delta=1$ for $k$-th stratum, follows a logistic regression model 
$\text{logit}\{r_k(W,\psi)\}=\psi_1+\psi_2 Z_1+\psi_3 A$. With $\psi=(1.5,-1,-0.5)$, we have 
about 45\% missingness for $Z_1=1$ and about 20\% missingness for $Z_1=0$. 
Since only two causes are considered in this simulation study, we posit a logistic regression 
model $\text{logit}\{ \rho_{k2}(W,\varphi) \}=\varphi_1+\varphi_2 Z_1+\varphi_3 A$  for $\rho_{k2}(W)$, 
the probability $P(V=2|\delta=1,W)$ in the $k$-th stratum. The parameter $\rho_{k1}(W)$ is estimated by $1-\rho_{k2}(W,\hat{\varphi})$.

We conducted simulations with sample size $n=1200$ and with different sets of values for $\theta_{kj}$, 
$\alpha_j$ and $\varphi_j$, $ j=1,2, k=1,2,3$.
We choose $(\theta_{11},\theta_{12})=(0.2,0.2),(\theta_{21},\theta_{22})=(0.5,0.5),(\theta_{31},\theta_{32})=(1,1)$, and $(\gamma_1,\gamma_2)=(1,1)$.
The following parameter settings of $\alpha_j=\log(1-VE_j)$ are considered for testing $H_{A0}$, $H_{Aj0}$, and $H_{B0}$ against the alternative hypotheses defined
 in Section \ref{testing},
$j=1,2$, where $c_0=\log(1-VE_0)=\log(1-0.3)=-0.3567$:
\begin{itemize}
\item[(1)] For testing  $H_{A0}$ and $H_{Aj0}$, $j=1,2$,
$M_1$:   $(\alpha_1,\alpha_2)=(\log(1-0.3),\log(1-0.3))$,
$M_2$:   $(\alpha_1,\alpha_2)=(\log(1-0.5),\log(1-0.3))$, and 
$M_3$:   $(\alpha_1,\alpha_2)=(\log(1-0.6),\log(1-0.3))$; 
\item[(2)] For testing $H_{B0}$, 
$N_1$:   $(\alpha_1,\alpha_2)=(\log(1-0.5),\log(1-0.5))$,
$N_2$:   $(\alpha_1,\alpha_2)=(\log(1-0.7),\log(1-0.5))$, and 
$N_3$:   $(\alpha_1,\alpha_2)=(\log(1-0.9),\log(1-0.5))$. 
\end{itemize}

The estimation procedures are examined under the setting $M_3$ of model (\ref{simu-model1}). IPW and AIPW estimators are compared with the complete-case (CC) estimator, 
which is obtained by solving (\ref{score}) based on the complete data only. Tables \ref{simu-M3-alpha} and \ref{simu-M3-VE}  show biases, the sample standard errors (SSE), the mean of the estimated standard errors (ESE), and 95\% empirical coverage probabilities (CP) of the estimators of $\alpha_1$, $\alpha_2$,$VE_1$, $VE_2$, and $VD(2,1)$  under the setting $M_3$ of model (\ref{simu-model1}) for $n=1200$ based on 1000 simulations. Both IPW and AIPW estimators have reasonably small bias when the model for $r_k(W)$ is correctly specified. The standard error estimators are fairly accurate, and the 95\% confidence intervals have reasonable coverage probabilities. The AIPW estimators are more efficient than IPW estimators, achieving more efficiency gain as the association between auxiliary $A$ and cause $V$ strengthens.  

The observed sizes and powers of the tests are examined under the settings $M_1$ to $M_3$ for testing $H_{A0}$ and $H_{Aj0}$, $j=1,2$, and $N_1$ to $N_3$ for testing $H_{B0}$, where $M_1$ and $N_1$ are the null hypotheses under $H_{A0}$ and $H_{B0}$, respectively. 
Tables \ref{table-testA}, \ref{table-testAj}, and \ref{table-testB} report empirical sizes and powers of the test statistics $\{U_{1}, U_{2}\}$ for testing $H_{A0}$,  the test statistics $\{U_{1j}, U_{2j}\}$ for testing $H_{Aj0}$, $j=1,2$, and  the test statistics $\{T_{1}, T_{2}\}$ for testing $H_{B0}$ under model (\ref{simu-model1})  for $n=1200$ at nominal level 0.05 based on 1000 simulations.
The empirical levels from all tests are closer to the nominal level 0.05 for both IPW and AIPW methods. 
The powers increase as the extend of departure from the corresponding null hypothesis increases.
When the correlation between auxiliary $A$ and cause $V$ becomes stronger, powers using AIPW method increase and are slightly higher than those using IPW methods.
% This result is expected, as the IPW estimators are less efficient than AIPW estimators.
% For multiple comparisons, the concept of power is more complex. There are several ways to formulate the definition of power.

\section{An application to a pseudo dataset for the Moderna COVE vaccine efficacy trial  }
\label{application}

%{\colr Show the data analysis of the practice dataset using Approach 2.}

We apply the proposed methods to a pseudo dataset designed to approximate the Moderna COVE vaccine efficacy 
trial of the mRNA-1273 vaccine (Baden et al., 2020).\nocite{Badenetal2020}  
The primary endpoint is virologically confirmed COVID-19 disease 
(e.g., Krause et al., 2020, Lancet; Mehrotra et al., 2020, Ann Int Med).\nocite{Krauseetal2020,mehrotra2020clinical}  
The data set approximately fits the Moderna COVE trial design, in terms of numbers of enrolled study participants,
numbers of COVID-19 endpoints in the two treatment groups, and through analysis of 1122
randomly sampled real SARS-CoV-2 Spike protein sequences downloaded from GISAID to determine the strain type (i.e. cause) $V$ 
with a realistic distribution of interest.

The COVE trial randomized adults at risk for COVID-19 to vaccine or placebo in one-to-one allocation
(administered at Day 1 and Day 29), and was designed to follow
participants for occurrence of the COVID-19 primary endpoint for 2 years.
Participants were enrolled and followed starting on July 27, 2020, and in late December of 2020
the U.S. FDA granted Emergency Use Authorization to the vaccine based on its demonstrated high vaccine efficacy.
Shortly after that, a process began to unblind study participants
and to offer the vaccine to placebo recipients. 
Our analysis restricts to the primary period of follow-up (pre-unblinding) and 
to the participants who tested negative for SARS-CoV-2 at enrollment and who received both 
vaccinations without specified protocol violations (the primary analysis cohort). 
Following the protocol-specified primary analysis COVID-19 endpoints are counted
starting 14 days post dose 2; we use as time origin 13 days post dose 2.

The data set includes 13,271 participants in the vaccine group with 72 COVID endpoint cases 
and 13,299 participants in the placebo group with 713 COVID endpoint cases. 
 To mimick the sequences expected from placebo arm COVID-19 endpoint
cases in the COVE trial, the sample of 1122 Spike protein sequences from GISAID was drawn from sequences
with deposition date between 
September 8 2020 and February 1 2021 (the approximate period of primary endpoint occurrence during 
blinded follow-up in the COVE trial) and with location the city of a COVE trial study site.  
 Based on the 1122 sampled GISAID sequences, 
the vast majority of amino acid positions (of 1273 positions in Spike) have more than 99\% of 
sequences matching the WA strain residue; 
for these positions there is not enough variability to support sieve analysis.  
However, for 7 amino acid positions, between 35 and 50 sequences (about 3 to 5\%) have a
 WA-strain mismatched residue; moreover one variant of concern (B.1.429, the ``California strain") has prevalence 
3.2\%.  Based on these data, for placebo arm endpoint cases
we generate $V$ from a Bernoulli random variable with 
success probability 0.04, which represents evaluation in sieve analysis of one of the 7 amino acid positions
 ($V=0$ is the WA strain residue, $V=1$ 
is the non-WA strain residue), or of the B.1.429 variant ($V=0$ is the WA strain, $V=1$ is the B.1.429 variant).  
To create a sieve effect where VE is less against the $V=1$ genotype, for 
vaccine recipients we draw $V$ from a Bernoulli distribution
with success probability 0.10. 
 %The sample size of 785 sequences was chosen as an estimate of the number of sequences that may be 
%available for the final blinded-period of follow-up sieve analysis (pooled over vaccine and placebo).

%The data set includes a binary mark/cause variable $V$, the indicator of vaccine-mismatch at a 
%particular amino acid position or haplotype set of positions. 
$V$ was measured from 25 (34.7\%) of the 72 vaccine recipient cases and from 382 (53.6\%) of the 
713 placebo recipient cases.
Because the SARS-CoV-2 viral load ($VL$) is correlated with the probability of missingness and the strain type $V$, 
we consider it as an auxiliary variable in the analysis.
%The SARS-CoV-2 viral load is measured in COVID endpoint cases from the same blood sample used for measuring the SARS-CoV-2 sequence.
A special problem that needs to be addressed is that samples with low viral load may have the probability of missingness very close to 1. 
This ``positivity problem" can make methods that use inverse probability weighting perform unstably. To allow our IPW and AIPW methods to work more robustly, we classify the 
COVID-19 endpoint participants with low viral load as a minor type of failure causes while the study endpoint of interest is still COVID-19. Specifically, $V$ is redefined as the cause of failure with three types:  $1=$ vaccine-matched genotype AND viral load above minimum threshold; $2=$ vaccine-mismatched genotype AND viral load above minimum threshold; and $3=$ viral load below minimum threshold.
%\begin{itemize}
%\item $V=1$: vaccine-matched genotype AND viral load above minimum threshold;
%\item $V=2$: vaccine-mismatched genotype AND viral load above minimum threshold;
%\item $V=3$: viral load below minimum threshold.
%\end{itemize}
Here, the minimum threshold is chosen large enough such that based on an empirical analysis
COVID-19 endpoint cases with $VL$ at the minimum threshold have at least 0.05--0.10 probability that
the sequence genotype is observed.  We use the minimum threshold $h_0=1$ for illustration.

Since the cause $V=3$ can be determined based on the observed viral load $<h_0$ (an auxiliary variable), the probability of it being not missing is one
 conditional on the observed viral load. The missing at random (MAR) assumption still holds:
\begin{eqnarray}
&&P(R=1|\delta=1,T,Z,VL,h_0,V) \nonumber \\
& &= P(R=1|\delta=1,T,Z,VL,h_0)\nonumber\\
 && = I(VL<h_0) + P(R=1|\delta=1,T,Z,VL,VL\ge h_0)I(VL\ge h_0).\nonumber
\end{eqnarray}

The data analysis uses $K=3$ baseline strata defined by geography and calendar time. 
Let $T$ be the time from 13 days post dose 2 until the COVID-19 endpoint. 
We consider the following stratified cause-specific proportional hazards model:
\begin{eqnarray}
\label{data-model1}
 \lambda_{kj}(t|z)= \lambda_{k0}(t)\exp\big (\alpha_{j}\text{Trt}+\gamma_{1j}\text{Highrisk} +\gamma_{2j} \text{Age}_{65}+\gamma_{3j} \text{Minority} +\gamma_{4j} \text{Sex}\big ),
\end{eqnarray}
for $j=1,2,3$ and $k=1,2,3$, where Trt is the vaccine group indicator, 
Highrisk is the baseline covariate high risk/at-risk pre-existing condition (1=yes, 0=no), 
$\text{Age}_{65}$ is the age group at enrollment (1=``65+", 0=``18-64"), 
Minority is the baseline covariate underrepresented minority status (1=minority, 0=non-minority), and   
Sex is sex assigned at birth (1=female, 0=male).

Let $Z=(\text{Trt},\text{Minority},\text{Highrisk},\text{Sex},\text{Age}_{65})$. 
We fit a logistic regression model with predictors $(1, \text{Trt}, VL)$ to estimate the conditional 
probability $P(R=1|\delta=1,T,Z,VL,VL\ge h_0)$ for each stratum. 
%When the observed viral load $<h_0$, the cause is defined as $V=2$, i.e. 
Note that $P(V=3|\delta=1,T,Z,VL<h_0)=1$ and 
$P(V=2|\delta=1,T,Z,VL<h_0)=P(V=1|\delta=1,T,Z,VL<h_0)=0$. Therefore, 
to implement the AIPW method, we only need to estimate  $P(V=2|\delta=1, T,Z,VL,VL\ge h_0)$. It is modeled using logistic regression with predictors $(1, T, \text{Trt}, VL)$. 
Then, the estimates of 
$P(V=1|\delta=1, T,Z,VL,VL\ge h_0)$ can be obtained through the relationship
 $P(V=1|\delta=1, T,Z,VL,VL\ge h_0)=1-P(V=2|\delta=1, T,Z,VL,VL\ge h_0)$.

Tables \ref{data-coef}-\ref{data-test} report the results for the estimation of covariate effects, 
the estimation of strain-specific vaccine efficacies, and the hypothesis testing for vaccine efficacies.
The analysis show that the vaccine is highly protective against vaccine-matched genotype ($V$=1) with point estimates of VE beyond 90\%, but  is not effective in protecting against infection when circulating strains are not well matched the vaccine strain.
The vaccine efficacy against vaccine-matched genotype is statistically significantly greater than the null level 30\% ($p$-value$<0.001$) and greater than the VE against vaccine-mismatched genotype ($p$-value$<0.001$).
The IPW and AIPW methods provide similar results while AIPW estimates are more efficient in terms of estimated standard errors.

To better illustrate our approach for $J>2$, we perform an additional analysis of this pseudo dataset using another mark variable, the Hamming distance to the vaccine-insert in the Spike protein. The Hamming distance is a count variable, which is the number of differing amino acids between the vaccine insert and the circulating Spike sequence. Some of its values appear less infrequently in the data which may cause an identifiability problem. Thus, we group Hamming distances into four classes: $0; \{1,2,3,4\}; \{5,6,7,8\}$; and greater than 8. Further considering the low viral load group,  we define failure causes $V^*$ as a categorical variable with five levels: $1=$ Hamming distance $=0$ AND $VL\ge h_0$; $2=$  Hamming distance $\in \{1,2,3,4\}$ AND $VL\ge h_0$; $3=$ Hamming distance $\in \{5,6,7,8\}$ AND $VL\ge h_0$; $4=$ Hamming distance $\ge 9$ AND $VL\ge h_0$; $5=$ $VL< h_0$.
%\begin{itemize}
%\item $V^*=1$: Hamming distance $=0$ AND $VL\ge h_0$;
%\item $V^*=2$: Hamming distance $\in \{1,2,3,4\}$ AND $VL\ge h_0$;
%\item $V^*=3$: Hamming distance $\in \{5,6,7,8\}$ AND $VL\ge h_0$;
%\item $V^*=4$: Hamming distance $\ge 9$ AND $VL\ge h_0$;
%\item $V^*=5$: $VL< h_0$.
%\end{itemize}
Results of the analysis using hamming distance are summarized in Tables \ref{data-coef2}-\ref{data-test2}. 
The vaccine efficacy is significantly greater than the null level 30\% for hamming distances less than 9. 
We also confirm a trend that the vaccine provides better protection against circulating viruses with smaller hamming distances.

%\section{Concluding Remarks}
%\label{conclusion}
%
%{\colr To be added later...}

\bigskip

%\noindent{\Large\bf Acknowledgements}
\section*{Acknowledgements}

This research was partially supported by  NIAID NIH award number R37AI054165.
Dr. Sun's research was also partially supported by the National Science Foundation grants DMS1915829.  

\renewcommand{\theequation}{A.\arabic{equation}}
\setcounter{equation}{0}

\section*{Appendix}

Let ${\cal F}_{t}=\sigma\{I(X_{ki}\le s, \delta_{ki}=1),$
$I(X_{ki}\le s, \delta_{ki}=0), V_{ki}$ $I(X_{ki}\le s, \delta_{ki}=1),
Z_{ki}(s); 0\le s\le t, i=1,\ldots,n_k, k=1,\ldots, K\}$ be the
(right-continuous) filtration generated by the full data processes
$\{N_{kij}(s), Y_{ki}(s),$  $Z_{ki}(s); 0\le s\le t, j=1,\ldots,J,
i=1,\ldots,n_k, k=1,\ldots, K\}$. 
Assume
$E(N_{kij}(dt)|{\cal F}_{t-})=E(N_{kij}(dt)|Y_{ki}(t),Z_{ki}(t))$,
that is, the cause-specific instantaneous failure rate at time $t$ given the observed
information up to time $t$ only depends on the failure status and the current covariate value. 
%Hence,
Under model (\ref{causeM1}),
the cause-specific intensity of $N_{kij}(t)$ with respect to ${\cal F}_t$ equals $Y_{ki}(t)\lambda_{kj}(t|Z_{ki}(t))$. 
Let $M_{kij}(t)=\int_0^t [N_{kij}(ds)-Y_{ki}(s)\lambda_{kj}(s|Z_{ki}(s))\,ds]$.
By Aalen and Johansen (1978), for $j\neq j^\prime$, $M_{kij}(\cdot)$ and
$M_{kij^\prime}(\cdot)$ are orthogonal square integrable martingales with respect to ${\cal F}_t$ for  
$j, j^\prime=1,\ldots,J$.

Let ${\cal F}_t^*={\cal F}_t \cup \{R_{ki},\delta_{ki} A_{ki};\; i=1,\ldots,n_k, k=1,\ldots, K\}$
be the right continuous filtration obtained by adding $R_{ki}$ and $\delta_{ki}A_{ki}$ to ${\cal F}_t$.
Let $Y_{ki}(t) \lambda_{kij}^*(t)$ be the intensity of $N_{kij}(t)$ with respect to ${\cal F}_t^*$
%$=P(T_{ki}=t,V_{ki}=j |X_{ki}\ge t, Z_{ki}(t), R_{ki},\delta_{ki}A_{ki})$. 
Then
%\begin{equation}
%\label{EQR}
$E(N_{kij}(dt)|{\cal F}_{t-}^*)=Y_{ki}(t)\lambda_{kij}^*(t)\,dt$.
%\end{equation}
%where $Y_{ki}(t)\lambda_{kij}^*(t)$ is the intensity of $N_{kij}(t)$ with respect to ${\cal F}_t^*$.
Assume that $\lambda_{kij}^*(t)$ is continuous in $t$.
Let $M_{kij}^*(t)=N_{kij}(t)-\int_0^t Y_{ki}(s) \lambda_{kij}^*(s)\,ds.$
By Aalen \& Johansen (1978),   the processes $M_{kij}^*(t)$ and
$M_{kij^\prime}^*(\cdot)$, $0\le t \le \tau$ are orthogonal square integrable martingales for $j\neq j^\prime$.

The following regularity conditions are assumed throughout the
rest of the paper. Most of the notation can be found at the end
of Section \ref{Preliminaries}.

\medskip
\noindent  {\bf Condition A}
\begin{itemize}

\item[(A.1)] $\ib_j$ has componentwise continuous second derivatives on $[0,1]$.
For each $k=1,\ldots, K$, the second partial derivative of
$\lambda_{0k}(t,v)$ with respect to $v$ exists and is continuous on
$[0,\tau]\times [0,1]$. The covariate process $Z_{ki}(t)$ has paths
that are left continuous and of bounded variation, and satisfies the
moment condition $E[\|Z_{ki}(t)\|^4\exp(2M\|Z_{ki}(t)\|)]<\infty$, where
$M$ is a constant such that $(v,\ib_j)\in [0,1]\times(-M,M)^p$
for all $v$ and $\|A\|=\max_{k,l} |a_{kl}|$ for a matrix $A=(a_{kl})$.

\item[(A.2)]  Each component of
$s_k^{(j)}(t,\theta)$ is continuous on $[0,\tau]\times[-M,M]^p$,
$\tilde s_k^{(j)}(t,\theta,\psi_k)$  is continuous on
$[0,\tau]\times[-M,M]^p\times [-L,L]^q$  for some $M, L>0$ and
$j=0,1,2$.
 $\sup_{t\in [0,\tau], \theta\in [-M,M]^p}$ $\|S_k^{(j)}(t, \theta)-s_k^{(j)}(t,\theta)\| =O_p(n^{-1/2})$, and
 $\sup_{t\in [0,\tau], \theta\in [-M,M]^p, \psi_k\in [-L,L]^q}$
 $\|\tilde S_k^{(j)}(t, \theta,\psi_k)-\tilde s_k^{(j)}(t,\theta,\psi_k)\|
 =O_p(n^{-1/2})$.

\item[(A.3)] The limit $p_k=\lim_{n\to\infty} n_k/n$ exists and $0<p_k<1$.
 $s_k^{(0)}(t,\theta)>0$ on $[0,\tau]\times [-M,M]^p$ and
the matrix $\Sigma_j=\sum_{k=1}^K p_k\Sigma_{kj}$  is positive
definite, where $\Sigma_{kj}=\sum_{k=1}^K \int_0^\tau I_k(t,\ib_j)$ $\lambda_{0kj}(t)s_k^{(0)}(t,\ib_j)\,dt$.

\item[(A.4)] The kernel function $K(\cdot)$ is symmetric with support $[-1,1]$ and has bounded variation.
 The bandwidth satisfies   $nh^2\to\infty$ and $nh^5=O(1)$ as $n\to\infty$.

\item[ (A.5)]  There is a $\varepsilon > 0$ such that $r_k(W_{ki}) \ge \varepsilon$ for all $k,i$ with $\delta_{ki}=1$.

\end{itemize}

Discussion of some of these conditions can be found in Sun {\sl et al.}\ (2009).

\bigskip

\comment{
\newcommand{\new}{\setlength{\parindent}{0pt}\vspace{10.0pt}
\setlength{\hangindent}{1.6pc}\setlength{\hangafter}{1}}

\bigskip
\medskip

\noindent{\Large\bf References}

\medskip

\new Dabrowska, D.M. (1997). Smoothed Cox regression. {\it The Annals
of Statistics} {\bf 25}, 1510--1540.

\new Gilbert, B. P. (2000). Comparison of competing risks failure
time methods and time-independent methods for assessing strain
variations in vaccine protection. \textit{Biometrics},
\textbf{19}, 3065--3086.

\new Holland, B. and Copenhaver, M. (1987). An Improved Sequentially Rejective Bonferroni Test Procedure. \textit{Biometrics}, 43(2), 417-423.

\new Horvitz, D. and Thompson, D. (1952). A generalization of sampling without replacement from a finite universe.  \textit{Journal of the American Statistical Association}, 47:663--685.

\new Kalbfleisch, J. D. and Prentice, R. L. (1980). \textit{The
Statistical Analysis of Failure Time Data.} Wiley, New York.

%\new Prentice, R. L., Kalbfleisch, J. D., Perterson, A. V.,
%Flournoy, N., Farewell, V. T. and Breslow, N. E. (1978). The
%analysis of failure times in the presence of competing risks.
%\textit{Biometrics} \textbf{34}, 541--554.
 
\new Robins, J., Rotnitzky, A., and Zhao, L. (1994). Estimation of regression coefficients when some regressors are not always observed.  \textit{Journal of the American Statistical Association,} 89:846--866.

\new Rubin, D. B. (1976). Inference and missing data.  \textit{Biometrika}, 63(3):581--592.}

\renewcommand{\bibname}{References}
\bibliography{misscause-COVID}

\newpage

\begin{table} [ht]
\tabcolsep 5pt
\caption{Biases, sample standard errors (SSE), mean of the estiamted standard errors (ESE), and 95\% empirical coverage probabilities (CP) of the estimators of $\alpha_1$ and $\alpha_2$  under the setting $M_3$ of model (\ref{simu-model1}) for $n=1200$ based on 1000 simulations.
\label{simu-M3-alpha}}
\begin{center}
\def\arraystretch{1.2}
\makeatletter
\def\hlinew#1{
  \noalign{\ifnum0=`}\fi\hrule \@height #1 \futurelet
   \reserved@a\@xhline}
 \begin{tabular}{c c c c c c c c c c}
 \hlinew{1pt}
  & \multicolumn{4}{c}{$\alpha_1$} && \multicolumn{4}{c}{$\alpha_2$}   \\
  \cline{2-5} \cline{7-10}
Method & bias & SSE & ESE & CP && bias & SSE & ESE & CP   \\
\cline{1-10}
& \multicolumn{9}{c}{Auxiliary association level setting (Aux0): Kendall's tau $=0$}\\
CC   & -0.2609	&0.1641	&0.1599	&0.639 &&-0.2621	&0.1341	&0.1326	&0.501 \\
IPW  & -0.0099	&0.1563	&0.1507	&0.941 &&-0.0130	&0.1218	&0.1218	&0.949 \\
AIPW & -0.0102	&0.1536	&0.1473	&0.938 &&-0.0120	&0.1157	&0.1172	&0.959 \\
\cline{1-10}                         
& \multicolumn{9}{c}{Auxiliary association level setting (Aux1): Kendall's tau $=0.3$}\\
CC   & -0.2631	&0.1655	&0.1608	&0.635 &&-0.2922	&0.1371	&0.1363	&0.421 \\
IPW  & -0.0092	&0.1560	&0.1516	&0.946 &&-0.0130	&0.1231	&0.1235	&0.952 \\
AIPW & -0.0099	&0.1496	&0.1455	&0.945 &&-0.0114	&0.1150	&0.1164	&0.960 \\
\cline{1-10}                         
& \multicolumn{9}{c}{Auxiliary association level setting (Aux2): Kendall's tau $=0.6$}\\
CC   & -0.2668	&0.1666	&0.1620	&0.621 &&-0.3411	&0.1429	&0.1428	&0.324 \\
IPW  & -0.0088	&0.1565	&0.1526	&0.945 &&-0.0137	&0.1249	&0.1264	&0.955 \\
AIPW & -0.0084	&0.1377	&0.1343	&0.947 &&-0.0111	&0.1101	&0.1109	&0.955 \\
\hlinew{1pt}
 \end{tabular}
\end{center}
\end{table}

\begin{table} [ht]
\tabcolsep 5pt
\caption{Biases, sample standard errors (SSE), mean of the estiamted standard errors (ESE), and 95\% empirical coverage probabilities (CP) of the estimators of $VE_1$, $VE_2$, and $VD(2,1)$ under the setting $M_3$ of model (\ref{simu-model1}) based on 1000 simulations.
\label{simu-M3-VE}}
\begin{center}
\def\arraystretch{1.2}
\makeatletter
\def\hlinew#1{
  \noalign{\ifnum0=`}\fi\hrule \@height #1 \futurelet
   \reserved@a\@xhline}
 \begin{tabular}{c c c c c c c c c c}
 \hlinew{1pt}
  & \multicolumn{4}{c}{IPW} && \multicolumn{4}{c}{AIPW}   \\
  \cline{2-5} \cline{7-10}
 & bias & SSE & ESE & CP && bias & SSE & ESE & CP \\
\cline{1-10}
& \multicolumn{9}{c}{Auxiliary association level setting (Aux0): Kendall's tau $=0$}\\
$VE_1$   & -0.0009	&0.0628	&0.0601	&0.941 &&-0.0006	&0.0614	&0.0587	&0.938  \\
$VE_2$   & 0.0039	&0.0848	&0.0847	&0.949 &&0.0037	&0.0806	&0.0815	&0.959  \\
$VD(2,1)$& 0.0336	&0.3785	&0.3694	&0.943 &&0.0362	&0.3814	&0.3708	&0.943  \\
\cline{1-10}                         
& \multicolumn{9}{c}{Auxiliary association level setting (Aux1): Kendall's tau $=0.3$}\\
$VE_1$   & -0.0012	&0.0629	&0.0605	&0.946 &&-0.0005	&0.0599	&0.0580	&0.945  \\
$VE_2$   & 0.0038	&0.0859	&0.0858	&0.952 &&0.0033	&0.0802	&0.0810	&0.960  \\
$VD(2,1)$& 0.0322	&0.3772	&0.3734	&0.953 &&0.0345	&0.3693	&0.3644	&0.947  \\
\cline{1-10}                         
& \multicolumn{9}{c}{Auxiliary association level setting (Aux2): Kendall's tau $=0.6$}\\
$VE_1$   & -0.0014	&0.0630	&0.0610	&0.945 &&-0.0004	&0.0549	&0.0536	&0.947  \\
$VE_2$   & 0.0041	&0.0872	&0.0878	&0.955 &&0.0035	&0.0768	&0.0771	&0.955  \\
$VD(2,1)$& 0.0309	&0.3806	&0.3795	&0.953 &&0.0249	&0.3282	&0.3243	&0.946  \\
\hlinew{1pt}
 \end{tabular}
\end{center}
\end{table}

\begin{table} [ht]
\tabcolsep 5pt
\caption{ Empirical sizes and powers of the test statistics $U_{1}$ and $U_{2}$ for testing $H_{A0}$ under model (\ref{simu-model1})  for $n=1200$ at nominal level 0.05 based on 1000 simulations.
\label{table-testA} }
\begin{center}
\def\arraystretch{1.2}
\makeatletter
\def\hlinew#1{
  \noalign{\ifnum0=`}\fi\hrule \@height #1 \futurelet
   \reserved@a\@xhline}
 \begin{tabular}{c c c c c c c}
 \hlinew{1pt}
  &  & \multicolumn{2}{c}{IPW}  && \multicolumn{2}{c}{AIPW}   \\
  \cline{3-4} \cline{6-7}
Model & Test  & $U_{1}$&$U_{2}$  && $U_{1}$&$U_{2}$    \\
\cline{1-7}
& & \multicolumn{5}{c}{(Aux0): Kendall's tau $=0$}\\
$M_{1}$ &Size       &0.053	&0.059&&0.055	&0.049  \\
$M_{2}$ &Power      &0.718	&0.584&&0.726	&0.600  \\
$M_{3}$ &           &0.973	&0.943&&0.980	&0.954  \\
\cline{1-7}                     
& & \multicolumn{5}{c}{(Aux1): Kendall's tau $=0.3$}\\
$M_{1}$ &Size       &0.051	&0.059&&0.055	&0.052  \\
$M_{2}$ &Power      &0.706	&0.576&&0.733	&0.606  \\
$M_{3}$ &           &0.972	&0.942&&0.981	&0.958  \\
\cline{1-7}                      
& & \multicolumn{5}{c}{(Aux2): Kendall's tau $=0.6$}\\
$M_{1}$ &Size       &0.055	&0.058&&0.049	&0.044  \\
$M_{2}$ &Power      &0.694	&0.562&&0.770	&0.672  \\
$M_{3}$ &           &0.971	&0.927&&0.994	&0.979  \\
\hlinew{1pt}
 \end{tabular}
\end{center}
\end{table}

\begin{table} [ht]
\tabcolsep 5pt
\caption{ Empirical sizes and powers of the test statistics $U_{1j}$ and $U_{2j}$ for testing $H_{Aj0}$, $j=1,2$, under model (\ref{simu-model1})  for $n=1200$ at nominal level 0.05 based on 1000 simulations.
\label{table-testAj} }
\begin{center}
\def\arraystretch{1.2}
\makeatletter
\def\hlinew#1{
  \noalign{\ifnum0=`}\fi\hrule \@height #1 \futurelet
   \reserved@a\@xhline}
 \begin{tabular}{c c c c c c c c c c c}
 \hlinew{1pt}
  &  & \multicolumn{4}{c}{IPW}  && \multicolumn{4}{c}{AIPW}   \\
  \cline{3-6} \cline{8-11}
Model & Test  & $U_{11}$&$U_{21}$&$U_{12}$&$U_{22}$   && $U_{11}$&$U_{21}$&$U_{12}$&$U_{22}$    \\
\cline{1-11}
& & \multicolumn{9}{c}{(Aux0): Kendall's tau $=0$}\\
$M_{1}$ &Size       &0.051	&0.053	&0.046	&0.047&&0.047	&0.054	&0.048	&0.042  \\
$M_{2}$ &Power      &0.811	&0.711	&0.042	&0.037&&0.819	&0.722	&0.045	&0.048  \\
$M_{3}$ &           &0.987	&0.971	&0.059	&0.054&&0.991	&0.979	&0.064	&0.045  \\
\cline{1-11}                     
& & \multicolumn{9}{c}{(Aux1): Kendall's tau $=0.3$}\\
$M_{1}$ &Size       &0.045	&0.052	&0.047	&0.045&&0.056	&0.052	&0.046	&0.046  \\
$M_{2}$ &Power      &0.799	&0.700	&0.044	&0.047&&0.829	&0.726	&0.044	&0.057  \\
$M_{3}$ &           &0.986	&0.970	&0.062	&0.049&&0.991	&0.979	&0.059	&0.046  \\
\cline{1-11}                      
& & \multicolumn{9}{c}{(Aux2): Kendall's tau $=0.6$}\\
$M_{1}$ &Size       &0.047	&0.054	&0.042	&0.050&&0.059	&0.049	&0.046	&0.041  \\
$M_{2}$ &Power      &0.788	&0.678	&0.046	&0.045&&0.858	&0.765	&0.054	&0.052  \\
$M_{3}$ &           &0.987	&0.968	&0.063	&0.048&&0.996	&0.992	&0.061	&0.047  \\
\hlinew{1pt}
 \end{tabular}
\end{center}
\end{table}

\begin{table} [ht]
\tabcolsep 5pt
\caption{ Empirical sizes and powers of the test statistics $T_{1}$ and $T_{2}$ for testing $H_{B0}$ under model (\ref{simu-model1})  for $n=1200$ at nominal level 0.05 based on 1000 simulations.
\label{table-testB} }
\begin{center}
\def\arraystretch{1.2}
\makeatletter
\def\hlinew#1{
  \noalign{\ifnum0=`}\fi\hrule \@height #1 \futurelet
   \reserved@a\@xhline}
 \begin{tabular}{c c c c c c c}
 \hlinew{1pt}
  &  & \multicolumn{2}{c}{IPW}  && \multicolumn{2}{c}{AIPW}   \\
  \cline{3-4} \cline{6-7}
Model & Test  & $T_{1}$&$T_{2}$  && $T_{1}$&$T_{2}$    \\
\cline{1-7}
& & \multicolumn{5}{c}{(Aux0): Kendall's tau $=0$}\\
$N_{1}$ &Size       &0.047	&0.061&&0.048	&0.064  \\
$N_{2}$ &Power      &0.766	&0.664&&0.762	&0.663  \\
$N_{3}$ &           &1.000	&1.000&&1.000	&1.000  \\
\cline{1-7}                     
& & \multicolumn{5}{c}{(Aux1): Kendall's tau $=0.3$}\\
$N_{1}$ &Size       &0.047	&0.064&&0.051	&0.059  \\
$N_{2}$ &Power      &0.755	&0.647&&0.775	&0.671  \\
$N_{3}$ &           &1.000	&1.000&&1.000	&1.000  \\
\cline{1-7}                      
& & \multicolumn{5}{c}{(Aux2): Kendall's tau $=0.6$}\\
$N_{1}$ &Size       &0.047	&0.061&&0.051	&0.066  \\
$N_{2}$ &Power      &0.746	&0.638&&0.850	&0.764  \\
$N_{3}$ &           &1.000	&1.000&&1.000	&1.000  \\
\hlinew{1pt}
 \end{tabular}
\end{center}
\end{table}

\begin{table} [ht]
\tabcolsep 5pt
\caption{Estimation of covariate effects for the practice COVID-19 vaccine efficacy trial data set using IPW and AIPW methods for the cause $V$
\label{data-coef}}
\begin{center}
\def\arraystretch{1.2}
\makeatletter
\def\hlinew#1{
  \noalign{\ifnum0=`}\fi\hrule \@height #1 \futurelet
   \reserved@a\@xhline}
 \begin{tabular}{c c c c c c c c}
 \hlinew{1pt}
  & \multicolumn{3}{c}{IPW} && \multicolumn{3}{c}{AIPW}   \\
  \cline{2-4} \cline{6-8}
 & Est. & SE & $p$-value && Est. & SE & $p$-value  \\
\cline{1-8}
& \multicolumn{7}{c}{Cause $V=1$}\\
Trt      & -2.439	&0.269	&0.000  &&-2.461	&0.161	&0.000	 \\   
Highrisk & 2.175	&0.135	&0.000  &&2.003	&0.082	&0.000	 \\
Age65+   & 1.123	&0.115	&0.000  &&1.198	&0.076	&0.000	 \\ 
Minority & -0.159	&0.117	&0.175  &&-0.049	&0.080	&0.540	 \\
Female   & -0.163	&0.115	&0.158  &&-0.095	&0.076	&0.210	 \\
\cline{1-8}                         
& \multicolumn{7}{c}{Cause $V=2$}\\
Trt      & -0.115	&0.690	&0.868  &&-0.245	&0.656	&0.709	 \\    
Highrisk & 2.732	&0.631	&0.000  &&2.205	&0.546	&0.000	 \\
Age65+   & 1.590	&0.666	&0.017  &&2.006	&0.757	&0.008	 \\
Minority & 0.064	&0.670	&0.924  &&0.455	&0.602	&0.450	 \\
Female   & -0.054	&0.643	&0.933  &&0.086	&0.644	&0.894	 \\
\cline{1-8}                         
& \multicolumn{7}{c}{Cause $V=3$}\\
Trt      & -0.842	&0.533	&0.114  &&-0.841	&0.532	&0.114	 \\    
Highrisk & 2.765	&0.628	&0.000  &&2.769	&0.627	&0.000	 \\
Age65+   & 1.350	&0.459	&0.003  &&1.341	&0.460	&0.004	 \\
Minority & 0.674	&0.466	&0.148  &&0.643	&0.469	&0.171	 \\
Female   & 0.162	&0.485	&0.738  &&0.139	&0.486	&0.775	 \\
\hlinew{1pt}
 \end{tabular}
\end{center}

\footnotesize{Est., the estimate of covariate coefficients; SE, the estimated standard error of the estimators of covariate coefficients; $p$-value pertaining to testing no covariate effect.}
\end{table}

\begin{table} [ht]
\tabcolsep 5pt
\caption{Estimation of strain-specific vaccine efficacies for the practice COVID-19 VE trial data using IPW and AIPW methods for the cause $V$
\label{data-VE}}
\begin{center}
\def\arraystretch{1.2}
\makeatletter
\def\hlinew#1{
  \noalign{\ifnum0=`}\fi\hrule \@height #1 \futurelet
   \reserved@a\@xhline}
 \begin{tabular}{c c c c c c c c c c}
 \hlinew{1pt}
 &  &  &  &  & \multicolumn{2}{c}{$H_{Aj1}: VE_j>0.3$} && \multicolumn{2}{c}{$H_{Aj2}: VE_j\ne 0.3$}  \\
\cline{6-7}\cline{9-10}
 & Est. & SE & 95\% LL & 95\% UL & $U_{1j}$ & $p$-value && $U_{2j}$ & $p$-value  \\
\cline{1-10}
& \multicolumn{9}{c}{IPW}\\
$VE_1$ &0.913	&0.024	&0.852	&0.948	    &-7.737 &$<0.001$	 &&59.868&$<0.001$             \\
$VE_2$ &0.108	&0.615	&-2.445	&0.769	& 0.351&0.657       &&0.123&	0.701    \\
$VE_3$ &0.569	&0.230	&-0.225	&0.848	&-0.910&0.205       &&0.828&	0.383    \\
%VE(0) &0.913	&0.0235	&0.852	&0.948	    &-7.737&$<0.001$	 &&59.868&$<0.001$             \\
%VE(1) &0.108	&0.615	&-2.445	&0.769	& 0.351&0.657       &&0.123&	0.701    \\
%VE(2) &0.569	&0.230	&-0.225	&0.848	&-0.910&0.205       &&0.828&	0.383    \\
\cline{1-10}                         
& \multicolumn{9}{c}{AIPW}\\
$VE_1$ & 0.915	&0.014	&0.883	&0.938	    &-13.041 &$<0.001$	 &&170.060&$<0.001$              \\
$VE_2$ & 0.217	&0.514	&-1.834	&0.784	& 0.171&0.583       &&0.0292 &	0.863    \\
$VE_3$ & 0.569	&0.230	&-0.225	&0.848	    &-0.909&0.181       &&0.826 &	0.369        \\
%VE(0) & 0.915	&0.0138	&0.883	&0.938	    &-13.041 &$<0.001$	 &&170.0599&$<0.001$              \\
%VE(1) & 0.217	&0.514	&-1.834	&0.784	& 0.171&0.583       &&0.0292 &	0.863    \\
%VE(2) & 0.569	&0.230	&-0.225	&0.848	    &-0.909&0.181       &&0.826 &	0.369        \\
\hlinew{1pt}
 \end{tabular}
\end{center}

\footnotesize{
%$VE_j$, VE for strain type $V=j$, $j=1,2,3$; 
Est., the estimate of vaccine efficacies; SE, the estimated standard error of the estimators of VEs; 95\% LL and 95\% UL, lower limits (LL) and upper limits (UL) of 95\% confidence intervals of vaccine efficacies}%; $H_{Aj1}: VE(j)>0.3$; $H_{Aj2}: VE(j)\ne 0.3$}
\end{table}

\begin{table} [ht]
\tabcolsep 5pt
\caption{Estimation of VD for the practice COVID-19 vaccine efficacy trial data using IPW and AIPW methods for the cause $V$
\label{data-VD}}
\begin{center}
\def\arraystretch{1.2}
\makeatletter
\def\hlinew#1{
  \noalign{\ifnum0=`}\fi\hrule \@height #1 \futurelet
   \reserved@a\@xhline}
 \begin{tabular}{c c c c c }
\hlinew{1pt}
 & Est. & SE & 95\% LL & 95\% UL  \\
\cline{1-5}
& \multicolumn{4}{c}{IPW}\\
VD(2,1) & 10.216&7.607	&2.374	&43.967  \\
%VD(3,2) & 0.483	&0.423	&0.0871	&2.682   \\
VD(1,2) & 0.098	&0.073	&0.023	&0.421   \\
%VD(2,3) & 2.069	&1.808	&0.373	&11.477  \\
\cline{1-5}                         
& \multicolumn{4}{c}{AIPW}\\
VD(2,1) & 9.176	&6.678	&2.204	&38.210  \\
%VD(3,2) & 0.551	&0.466	&0.105	&2.887   \\
VD(1,2) & 0.109	&0.079	&0.026	&0.454   \\
%VD(2,3) & 1.815	&1.534	&0.346	&9.513   \\
\hlinew{1pt}
 \end{tabular}
\end{center}

\footnotesize{
%VD($i,j$)=(1-VE($i$))/(1-VE($j$)), measures how much greater the vaccine efficacy is against a strain $V=j$ virus than against a strain $V=i$ virus; 
Est., the estimate of VD; SE, the estimated standard error of the estimators of VD; 95\% LL and 95\% UL, lower limits (LL) and upper limits (UL) of 95\% confidence intervals for VD}
\end{table}

\begin{table} [ht]
\tabcolsep 5pt
\caption{Results of hypothesis tests for the practice COVID-19 vaccine efficacy trial data using IPW and AIPW methods for the cause $V$
%{\colr R codes need to be changed. The current version is for $j=0,1,2$. For instance, $H_{B1}$: VE(0)$\ge$VE(1)$\ge$VE(2) with strict inequality for some $j \in (0,1,2)$}
\label{data-test}}
\begin{center}
\def\arraystretch{1.2}
\makeatletter
\def\hlinew#1{
  \noalign{\ifnum0=`}\fi\hrule \@height #1 \futurelet
   \reserved@a\@xhline}
 \begin{tabular}{c c c c c c c c c c c c}
 \hlinew{1pt}
& \multicolumn{2}{c}{$H_{A1}$} && \multicolumn{2}{c}{$H_{A2}$} && \multicolumn{2}{c}{$H_{B1}$} && \multicolumn{2}{c}{$H_{B2}$}\\
\cline{2-3}\cline{5-6}\cline{8-9}\cline{11-12}
& $U_1$ & $p$-value && $U_2$ & $p$-value && $T_1$ & $p$-value && $T_2$ & $p$-value\\
\cline{1-12}
IPW  &-7.737&$<0.001$&&59.991&$<0.001$&&3.121&0.001&&9.740&0.002\\
AIPW&-13.041&$<0.001$&&170.089&$<0.001$&&3.046&0.001&&9.276&0.002\\
\hlinew{1pt}
 \end{tabular}
\end{center}

\footnotesize{$H_{A1}$: $VE_j \ge 0.3$ with strict inequality for some $j\in\{1,2\}$; $H_{A2}$: $VE_j \neq 0.3$ for some $j\in\{1,2\}$; $H_{B1}$: $VE_1>VE_2$; $H_{B2}$: $VE_1 \neq  VE_2$}
\end{table}

\begin{table} [ht]
\tabcolsep 5pt
\caption{Estimation of covariate effects for the practice COVID-19 vaccine efficacy trial data set using IPW and AIPW methods for the cause $V^*$
\label{data-coef2}}
\begin{center}
\def\arraystretch{0.7}
\makeatletter
\def\hlinew#1{
  \noalign{\ifnum0=`}\fi\hrule \@height #1 \futurelet
   \reserved@a\@xhline}
 \begin{tabular}{c c c c c c c c}
 \hlinew{1pt}
  & \multicolumn{3}{c}{IPW} && \multicolumn{3}{c}{AIPW}   \\
  \cline{2-4} \cline{6-8}
 & Est. & SE & $p$-value && Est. & SE & $p$-value  \\
\cline{1-8}
& \multicolumn{7}{c}{Cause $V^*=1$}\\
Trt      & -2.880	&0.377	&0.000&&-2.925	&0.255	&0.000	\\
Highrisk & 2.087	&0.155	&0.000&&1.976	&0.113	&0.000	\\
Age65+   & 1.339	&0.137	&0.000&&1.372	&0.107	&0.000	\\
Minority & -0.228	&0.141	&0.106&&-0.163	&0.111	&0.141	\\
Female   & -0.050	&0.136	&0.714&&-0.049	&0.108	&0.652	\\
\cline{1-8}
& \multicolumn{7}{c}{Cause $V^*=2$}\\
Trt      & -2.791	&0.638	&0.000&&-2.934	&0.730	&0.000  \\
Highrisk & 2.153	&0.341	&0.000&&1.908	&0.309	&0.000	\\
Age65+   & 1.108	&0.302	&0.000&&1.090	&0.309	&0.000	\\
Minority & 0.016	&0.326	&0.960&&0.162	&0.304	&0.595	\\
Female   & -0.256	&0.307	&0.406&&-0.049	&0.325	&0.880	\\
\cline{1-8}
& \multicolumn{7}{c}{Cause $V^*=3$}\\
Trt      & -1.493	&0.561	&0.008&&-1.513	&0.451	&0.001  \\
Highrisk & 2.373	&0.511	&0.000&&2.081	&0.372	&0.000	\\
Age65+   & 0.359	&0.388	&0.356&&0.643	&0.328	&0.050	\\
Minority & -0.063	&0.372	&0.866&&0.156	&0.321	&0.627	\\
Female   & -0.839	&0.388	&0.031&&-0.649	&0.332	&0.051	\\
\cline{1-8}
& \multicolumn{7}{c}{Cause $V^*=4$}\\
Trt      & -0.635	&0.502	&0.206&&-0.651	&0.402	&0.105  \\
Highrisk & 2.900	&0.475	&0.000&&2.301	&0.389	&0.000	\\
Age65+   & 0.581	&0.459	&0.205&&1.036	&0.372	&0.005	\\
Minority & 0.077	&0.425	&0.856&&0.443	&0.365	&0.225	\\
Female   & 0.099	&0.430	&0.817&&0.403	&0.409	&0.325	\\
\cline{1-8}
& \multicolumn{7}{c}{Cause $V^*=5$}\\
Trt      & -0.842	&0.533	&0.114&&-0.841	&0.533	&0.114	\\
Highrisk & 2.765	&0.628	&0.000&&2.769	&0.627	&0.000	\\
Age65+   & 1.350	&0.459	&0.003&&1.341	&0.460	&0.004	\\
Minority & 0.674	&0.466	&0.148&&0.643	&0.469	&0.171	\\
Female   & 0.162	&0.485	&0.738&&0.139	&0.486	&0.775	\\
\hlinew{1pt}
 \end{tabular}
\end{center}

\footnotesize{Est., the estimate of covariate coefficients; SE, the estimated standard error of the estimators of covariate coefficients; $p$-value pertaining to testing no covariate effect.}
\end{table}

\begin{table} [ht]
\tabcolsep 5pt
\caption{Estimation of strain-specific vaccine efficacies for the practice COVID-19 VE trial data using IPW and AIPW methods for the cause $V^*$
\label{data-VE2}}
\begin{center}
\def\arraystretch{1.2}
\makeatletter
\def\hlinew#1{
  \noalign{\ifnum0=`}\fi\hrule \@height #1 \futurelet
   \reserved@a\@xhline}
 \begin{tabular}{c c c c c c c c c c}
 \hlinew{1pt}
 &  &  &  &  & \multicolumn{2}{c}{$H_{Aj1}: VE_j>0.3$} && \multicolumn{2}{c}{$H_{Aj2}: VE_j\ne 0.3$}  \\
\cline{6-7}\cline{9-10}
 & Est. & SE & 95\% LL & 95\% UL & $U_{1j}$ & $p$-value && $U_{2j}$ & $p$-value  \\
\cline{1-10}
& \multicolumn{9}{c}{IPW}\\
$VE_1$ &0.944	&0.021	&0.882	&0.973	&-6.694	&0.000	&&44.804	&0.000     \\
$VE_2$ &0.939	&0.039	&0.786	&0.982	&-3.817	&0.000	&&14.572	&0.000     \\
$VE_3$ &0.775	&0.126	&0.325	&0.925	&-2.026	&0.019	&&4.105	&0.039         \\
$VE_4$ &0.470	&0.266	&-0.417	&0.802	&-0.554	&0.297	&&0.307	&0.573         \\
$VE_5$ &0.569	&0.230	&-0.225	&0.848	&-0.910	&0.170	&&0.828	&0.339         \\
\cline{1-10}
& \multicolumn{9}{c}{AIPW}\\
$VE_1$ &0.946	&0.014	&0.912	&0.967	&-10.061	&0.000	&&101.234	&0.000 \\
$VE_2$ &0.947	&0.039	&0.778	&0.987	&-3.530	&0.000	&&12.462	&0.001     \\
$VE_3$ &0.780	&0.099	&0.467	&0.909	&-2.564	&0.006	&&6.573	&0.011         \\
$VE_4$ &0.479	&0.210	&-0.147	&0.763	&-0.733	&0.227	&&0.537	&0.481         \\
$VE_5$ &0.569	&0.230	&-0.225	&0.848	&-0.909	&0.160	&&0.826	&0.357         \\
\hlinew{1pt}
 \end{tabular}
\end{center}

\footnotesize{
%VE($j$), VE for mark $V^*=j$, $j=0,1,2,3,4$; 
Est., the estimate of vaccine efficacies; SE, the estimated standard error of the estimators of VEs; 95\% LL and 95\% UL, lower limits (LL) and upper limits (UL) of 95\% confidence intervals of vaccine efficacies}%; $H_{Aj1}: VE(j)>0.3$; $H_{Aj2}: VE(j)\ne 0.3$}
\end{table}

\begin{table} [ht]
\tabcolsep 5pt
\caption{Estimation of VD for the practice COVID-19 vaccine efficacy trial data using IPW and AIPW methods for the cause $V^*$
\label{data-VD2}}
\begin{center}
\def\arraystretch{1.2}
\makeatletter
\def\hlinew#1{
  \noalign{\ifnum0=`}\fi\hrule \@height #1 \futurelet
   \reserved@a\@xhline}
 \begin{tabular}{c c c c c }
\hlinew{1pt}
 & Est. & SE & 95\% LL & 95\% UL  \\
\cline{1-5}
& \multicolumn{4}{c}{IPW}\\
VD(2,1) &1.092	&0.811	&0.255	&4.679 \\
VD(3,2) &3.664	&3.123	&0.689	&19.476\\
VD(4,3) &2.359	&1.788	&0.534	&10.418\\
%VD(5,4) &0.813	&0.595	&0.194	&3.416 \\
VD(1,2) &0.915	&0.679	&0.214	&3.921 \\
VD(2,3) &0.273	&0.233	&0.051	&1.451 \\
VD(3,4) &0.424	&0.321	&0.096	&1.873 \\
%VD(4,5) &1.230	&0.901	&0.293	&5.167 \\
\cline{1-5}
& \multicolumn{4}{c}{AIPW}\\
VD(2,1) &0.991	&0.818	&0.196	&5.000 \\
VD(3,2) &4.140	&3.680	&0.725	&23.638\\
VD(4,3) &2.367	&1.694	&0.582	&9.623 \\
%VD(5,4) &0.828	&0.552	&0.224	&3.060 \\
VD(1,2) &1.009	&0.833	&0.200	&5.090 \\
VD(2,3) &0.242	&0.215	&0.042	&1.379 \\
VD(3,4) &0.423	&0.302	&0.104	&1.718 \\
%VD(4,5) &1.209	&0.806	&0.327	&4.468 \\
\hlinew{1pt}
 \end{tabular}
\end{center}

\footnotesize{
%VD($j,k$)=(1-VE($j$))/(1-VE($k$)), measures how much greater the vaccine efficacy is against a strain $V=k$ virus than against a strain $V=j$ virus; 
Est., the estimate of VD; SE, the estimated standard error of the estimators of VD; 95\% LL and 95\% UL, lower limits (LL) and upper limits (UL) of 95\% confidence intervals for VD}
\end{table}

\begin{table} [ht]
\tabcolsep 5pt
\caption{Results of hypothesis tests for the practice COVID-19 vaccine efficacy trial data using IPW and AIPW methods for the cause $V^*$
\label{data-test2}}
\begin{center}
\def\arraystretch{1.2}
\makeatletter
\def\hlinew#1{
  \noalign{\ifnum0=`}\fi\hrule \@height #1 \futurelet
   \reserved@a\@xhline}
 \begin{tabular}{c c c c c c c c c c c c}
 \hlinew{1pt}
& \multicolumn{2}{c}{$H_{A1}$} && \multicolumn{2}{c}{$H_{A2}$} && \multicolumn{2}{c}{$H_{B1}$} && \multicolumn{2}{c}{$H_{B2}$}\\
\cline{2-3}\cline{5-6}\cline{8-9}\cline{11-12}
& $U_1$ & $p$-value && $U_2$ & $p$-value && $T_1$ & $p$-value && $T_2$ & $p$-value\\
\cline{1-12}
IPW  &-6.694&$<0.001$&&63.787&$<0.001$&&0.119&0.012&&3.616&0.283\\
AIPW&-10.061&$<0.001$&&120.806&$<0.001$&&-0.011&0.015&&4.003&0.236\\
\hlinew{1pt}
 \end{tabular}
\end{center}

\footnotesize{$H_{A1}$: $VE_j\ge 0.3$ with strict inequality for some $j\in\{1,2,3,4\}$; 
$H_{A2}$: $VE_j\neq 0.3$ for some $j\in\{1,2,3,4\}$; 
$H_{B1}$: $VE_1\ge VE_2 \ge VE_3 \ge VE_4$ with at least one strict inequality; 
$H_{B2}$: $VE_i \neq  VE_j$ for at least one pair of $\{(i,j)| i<j, i,j \in\{1,2,3,4\}\}$}
\end{table}

\end{document}